\documentclass[10pt, final, journal]{IEEEtran}

\usepackage[utf8]{inputenc}
\usepackage{tikz}
\usepackage{rotating}
\usepackage{amsmath,amssymb,amsthm}
\usepackage{listings}
\usepackage[textsize=tiny]{todonotes}
\usepackage{ wasysym }
\usepackage{pifont}
\usepackage{csvsimple}
\usepackage{siunitx}
\usepackage{tabularx}
\usepackage{xifthen}
\usepackage{subcaption}
\usepackage[T1]{fontenc}
\usepackage{dblfloatfix}
\usepackage{url}

\usepackage{xstring}
\newcommand{\optnum}[2]{\IfInteger{#1}{\ifnum0=0#1\relax0*\else\num{#1}\fi}{\IfDecimal{#1}{\num{#1}}{#2}}}
\usepackage{footnote}

\usepackage{tikz}
\usetikzlibrary{arrows}
\usetikzlibrary{arrows.meta}
\usetikzlibrary{positioning}  
\usetikzlibrary{shadows}
\usetikzlibrary{arrows,backgrounds,calc,chains,
matrix,positioning,shapes,shapes.geometric,decorations,
shapes.arrows, decorations.pathmorphing, decorations.pathreplacing, decorations.markings}

\newtheorem{example}{Example}

\usepackage{etoolbox}
\makeatletter
\patchcmd{\maketitle}{\@copyrightspace}{}{}{}
\makeatother

\usepackage{cite}
\usetikzlibrary{backgrounds,calc,chains,matrix,positioning,shapes,shapes.geometric,shapes.arrows}

\tikzset{%
	font={\footnotesize},
	vertex/.style={draw,circle,inner sep=0pt,minimum width=0.5cm,minimum height=0.5cm},
	zeroterm/.style={below,inner sep=0pt,font=\tiny}
}
\newcommand{\ket}[1]{\ensuremath{\left|#1\right\rangle}}

\IEEEoverridecommandlockouts

\title{An Efficient Methodology for Mapping \\Quantum Circuits to the IBM QX Architectures}

\author{Alwin Zulehner~\IEEEmembership{Student Member,~IEEE,} Alexandru Paler, and Robert Wille~\IEEEmembership{Senior Member,~IEEE}\\
	alwin.zulehner@jku.at\hspace{8mm}alexandru.paler@jku.at\hspace{8mm}robert.wille@jku.at}

\date{}

\usepackage[keeplastbox]{flushend}
\begin{document}

\maketitle

\begin{abstract}
In the past years, quantum computers more and more have evolved from an academic idea to an upcoming reality.
IBM's project \emph{IBM~Q} can be seen as evidence of this progress. Launched in March 2017 with the goal to provide access to quantum computers for a broad audience, 
this allowed users to conduct quantum experiments on a 5-qubit and, since June 2017, also on a 16-qubit quantum computer (called \emph{IBM~QX2} and \emph{IBM~QX3}, respectively). Revised versions of these 5-qubit and 16-qubit quantum computers (named \emph{IBM~QX4} and \emph{IBM~QX5}, respectively) are available since September 2017.
In order to use these, the desired quantum functionality (e.g.~provided in terms of a quantum circuit) has to be properly mapped
so that the underlying physical
constraints are satisfied -- a complex task. 
This demands solutions to automatically and efficiently conduct this mapping process. 

In this paper, we propose a methodology which addresses this problem, i.e.~maps the given quantum functionality to a realization
which satisfies all constraints given by the architecture and, at the same time, keeps the overhead in terms of additionally required quantum gates minimal. The proposed methodology is generic, can easily be configured for similar future architectures, and is fully integrated into IBM's SDK.
Experimental evaluations show that the proposed approach clearly outperforms IBM's own mapping solution. In fact, for many quantum circuits, the proposed approach determines a mapping to the IBM architecture within minutes, while IBM's solution suffers from long runtimes and runs into a timeout of 1 hour in several cases. As an additional benefit, the proposed approach yields mapped circuits with smaller costs (i.e.~fewer additional gates are required). All implementations of the proposed methodology is publicly available at~\url{http://iic.jku.at/eda/research/ibm_qx_mapping}.
\end{abstract}

\section{Introduction}
\label{sec:intro}

Quantum computers and quantum algorithms have received lots of interests in the past -- of course, mainly motivated by their ability 
to solve certain tasks significantly faster than classical algorithms~\cite{NC:2000,DBLP:journals/siamcomp/Shor97,DBLP:conf/stoc/Grover96,deutsch1992rapid}. These quantum algorithms are described by so-called quantum circuits, a sequence of gates that are applied to the qubits of a quantum computer. While theoretical algorithms have already been developed in the last century (e.g.~\cite{DBLP:journals/siamcomp/Shor97,DBLP:conf/stoc/Grover96,deutsch1992rapid}), physical realizations 
have been considered ``dreams of the future'' for a long time.
This changed in recent years in which quantum computers more and more evolved from an academic idea to an upcoming reality.

IBM's project \emph{IBM~Q}~\cite{ibmQ}, which launched in March 2017 with the goal to provide access to a quantum computer to the broad audience, can be seen as evidence of this progress. 
Initially, they started with the 5 qubit quantum processor \emph{IBM~QX2}, on which anyone could run experiments through cloud access. In June 2017, IBM added a 16 qubit quantum processor named \emph{IBM QX3} to their cloud~\cite{qxbackends} and, thus, more than tripled the number of available qubits within a few months. Since then, IBM has been working intensely on improving their quantum computers -- leading to 5-qubit and 16-qubit quantum computers (named \emph{IBM~QX4} and \emph{IBM~QX5}, respectively) which were added to the cloud in September 2017. 

The rapid progress in the number of available qubits is still going on. While IBM has already manufactured a 20-qubit quantum computer which is available for their partners and members of the \emph{IBM Q} network, as well as a prototype of a 50-qubit processor, other well-known companies like Google have also announced the intent to manufacture quantum chips with 49 qubits (using architectures as described in~\cite{neill2017blueprint}) in the near future to show quantum supremacy~\cite{boixo2016characterizing,courtland2017google}.

However, in order to use these physical realizations, the desired quantum functionality to be executed has to properly be mapped so that the underlying physical constraints are satisfied. This constitutes a complex task. 
One issue is that the desired functionality (usually described by higher level components) has to be decomposed into elementary operations supported by the \emph{IBM QX} architectures.
Furthermore, there exist physical limitations, namely that certain quantum operations can only be applied to selected physical qubits of the \emph{IBM~QX} architectures. Consequently, the logical qubits of a quantum circuit have to be mapped to the physical qubits of the quantum computer such that all 
operations can be conducted.
Since it is usually not possible to determine a mapping such that all constraints are satisfied throughout the whole circuit, this mapping may change over time. To this end, additional gates, e.g.~realizing SWAP operations, are inserted in order to ``move'' the logical qubits to other physical ones. 
They affect the reliability of the circuit (each further gate increases the potential for errors during the quantum computation)
as well as the execution time of the quantum algorithm. Hence, 
their number should be kept as small as possible.

While there exist several methods to address the first issue, i.e.~how to efficiently map higher level components to elementary operations (see~\cite{DBLP:journals/tcad/AmyMMR13,
	MWZ:2011,matsumoto2008representation,WSOD:2013}),
there is hardly any work on how to efficiently satisfy the additional constraints for these new and real architectures. 
Although there are similarities with recent work on nearest neighbor optimization of quantum circuits as proposed in~\cite{DBLP:journals/tcad/WilleLD14,DBLP:journals/qip/SaeediWD11,DBLP:conf/aspdac/WilleKWRCD16,SSP:2013,DBLP:conf/aspdac/ShafaeiSP14,DBLP:conf/rc/WilleQIYM16,zulehner2017exact}, 
they are not applicable since simplistic architectures with \mbox{1-dimensional} or 2-dimensional layouts are assumed in that work which have significantly less restrictions. 
Even IBM's own solution, which is provided by means of the Python SDK \emph{QISKit}~\cite{qiskit}
fails in many cases since the random search employed there does not cope with the underlying complexity and cannot generate a result in acceptable time.

The above motivates a solution that is as efficient as 
circuit designers e.g.~in the classical domain, take for granted today.
In this work\footnote{A preliminary version of this work is available at~\cite{DBLP:conf/date/ZulehnerPW18}.}, we propose a corresponding methodology.
To this end, a multi-step approach is introduced which utilizes a \mbox{depth-based} partitioning and $A^*$ as underlying search algorithm as well as further optimizations such as a \mbox{look-ahead} scheme and the ability to determine the initial mapping of the qubits throughout the mapping process (instead of fixing the initial mapping at the beginning of the algorithm). 
The resulting methodology is generic, i.e.~it can directly be applied to all existing QX~architectures as well as similar upcoming architectures which may come in the future (and~architectures whose constraints can be formulated in a similar way). 
Finally, we integrated the methodology into IBM's Python SDK \emph{QISKit} -- allowing for a more realistic performance evaluation since post-mapping optimizations provided by IBM are additionally considered. 

Experimental evaluations confirmed the benefits and allowed for an explicit analysis of the effects of the respective optimizations incorporated into the proposed methodology.  The results clearly show that the methodology is able to cope with the complexity of satisfying the constraints discussed above. Using this solution, QX-compatible mappings for many quantum circuits can be determined within minutes, 
while IBM's own solution suffers from long runtimes and runs into a timeout of 1 hour in these cases.
Moreover, as an additional benefit, realizations with smaller costs (i.e.~fewer additional gates) are obtained.
All implementations are publicly available at~\url{http://iic.jku.at/eda/research/ibm_qx_mapping} and, as mentioned above, have been integrated into IBM's own SDK -- resulting in an advanced and integrated mapping scheme for the QX architectures provided by IBM. 

This paper is structured as follows. In Section~\ref{sec:background}, we review quantum circuits as well as the \emph{IBM QX} architectures. In Section~\ref{sec:mapping}, we discuss 
the process to map a given quantum circuit to the \emph{IBM QX} architectures. How to particularly cope with the problem of satisfying the additional constraints is covered in Section~\ref{sec:proposed}. In Section~\ref{sec:results}, the performance of the proposed mapping scheme is analyzed and compared to the performance of the solution provided by IBM. Section~\ref{sec:conclusion} concludes the paper.

\section{Background}
\label{sec:background}

In this section, we briefly review the basics of quantum circuits and the \emph{IBM QX} architectures. 
\subsection{Quantum Circuits}
\label{sec:circuits}

Classical computations and circuits use bits as information units. In contrast, quantum circuits perform their computations on qubits~\cite{NC:2000}. These qubits can not only be in one of the two basis states $\ket{0}$ or $\ket{1}$, but also in a superposition of both -- allowing for the representation of all possible $2^n$ basis states of $n$ qubits concurrently. This so-called quantum parallelism serves as basis for algorithms that are significantly
faster on quantum computers than on classical machines.

To this end, the qubits of a quantum circuit are manipulated by quantum operations represented by so-called quantum gates. 
These operations can either operate on a single qubit, or on multiple ones. For multi-qubit gates, we distinguish target qubits and control qubits. The value of the target qubits is modified in the case that the control qubits are set to basis state $\ket{1}$.
The \emph{Clifford+T} library~\cite{DBLP:journals/tcad/AmyMMR13}, which is composed of the single-qubit gates \emph{H} (Hadamard gate) and \emph{T} (Phase shift by $\pi/4$), as well as the two-qubit gate \emph{CNOT} (controlled NOT), represents a universal set of quantum operations (i.e.~all quantum computations can be implemented by a circuit composed of gates from this library).

To describe quantum circuits, high level quantum languages (e.g.~Scaffold~\cite{abhari2012scaffold} or Quipper~\cite{DBLP:conf/pldi/GreenLRSV13}), quantum assembly languages (e.g.~OpenQASM~2.0 developed by IBM~\cite{cross2017open}), or circuit diagrams are employed. 
In the following, we use the latter to describe quantum circuits (but the proposed approach has also been applied using the other descriptions as well). In a circuit diagram,
qubits are represented by horizontal lines, which are passed through quantum gates. In contrast to classical circuits, this however does not describe a connection of wires with a physical gate, but defines (from left to right) in which order the quantum gates are applied to the qubits.

\begin{figure}
	\centering
\begin{tikzpicture}
\draw[line width=0.300000] (0.500000,1.250000) -- (4.5000000,1.250000);
\draw (0.400000,1.250000) node [left] {$q_0$};
\draw (4.600000,1.250000) node [right] {$q_0$};
\draw[line width=0.300000] (0.500000,0.750000) -- (4.5000000,0.750000);
\draw (0.400000,0.750000) node [left] {$q_1$};
\draw (4.600000,0.750000) node [right] {$q_1$};
\draw[line width=0.300000] (0.500000,0.250000) -- (4.5000000,0.250000);
\draw (0.400000,0.250000) node [left] {$q_2$};
\draw (4.600000,0.250000) node [right] {$q_2$};

\draw[line width=0.3,fill=white] (0.80,1.05) rectangle ++(0.4,0.4); \node at (1,1.25) {\sf H};

\draw[line width=0.300000] (1.750000,0.55000) -- (1.750000,1.250000);
\draw[fill] (1.750000,1.250000) circle (0.100000);
\draw[line width=0.3] (1.75, 0.75) circle (0.2);

\draw[line width=0.3,fill=white] (2.3,0.05) rectangle ++(0.4,0.4); \node at (2.5,0.25) {\sf T};

\draw[line width=0.300000] (3.250000,0.750000) -- (3.250000,0.050000);
\draw[fill] (3.250000,0.750000) circle (0.100000);
\draw[line width=0.3] (3.25, 0.25) circle (0.2);

\draw[line width=0.300000] (4.0000,1.250000) -- (4.0000,0.050000);
\draw[fill] (4.0000,1.250000) circle (0.100000);
\draw[line width=0.3] (4., 0.25) circle (0.2);
\end{tikzpicture}
	\caption{Circuit diagram of a quantum circuit}
	\label{fig:quantum_circuit}
\end{figure}
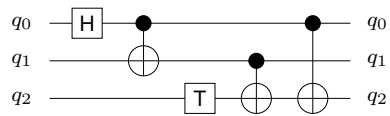

\begin{example}
	Fig.~\ref{fig:quantum_circuit} shows the circuit diagram of a quantum circuit. The quantum circuit is composed of three qubits and five gates.
The single-qubit gates \emph{H} and \emph{T} are represented by boxes labeled with H and T, respectively, while the control and target qubit of the CNOT gate are represented by $\bullet$ and~$\oplus$, respectively. 
	First, a Hadamard operation is applied to qubit~$q_0$. Then, a CNOT operation with target $q_1$ and control qubit $q_0$ is conducted -- followed by a T-gate that is applied to $q_2$. Finally, two more CNOTs are applied.
\end{example}

\subsection{IBM's QX Architectures}
\label{sec:architecture}

In this work, we consider how to efficiently map a quantum circuit to the \emph{IBM QX} architectures provided by the project \emph{IBM Q}~\cite{ibmQ}. 
IBM provides a Python SDK named \emph{QISKit}~\cite{qiskit} that allows a designer to describe quantum circuits, to simulate them, and to execute them on the real device (a \mbox{so-called} \emph{backend}) in their cloud. The first backend composed of 5 qubits and called \emph{IBM QX2} was launched in March 2017. In June 2017, IBM launched a second one called \emph{IBM QX3} which is composed of 16 physical qubits that are connected with coplanar waveguide bus resonators~\cite{qxbackends}. Quantum operations are conducted by applying microwave pulses to the qubits. In September 2017, IBM launched revised versions of their \mbox{5-qubit} and 16-qubit backends named \emph{IBM QX4} and \emph{IBM QX5}, respectively.

The \emph{IBM QX} architectures support the elementary single qubit operation $U(\theta, \phi	,\lambda)=R_z(\phi)R_y(\theta)R_z(\lambda)$ (i.e.~an Euler decomposition) that is composed by two rotations around the $z$-axis and one rotations around the $y$-axis, as well as the CNOT operation. 
By adjusting the parameters $\theta$, $\phi$, and $\lambda$, single-qubit operations of other gate libraries like the $H$ or the $T$ gate (cf.~Section~\ref{sec:circuits}) can be realized (among others like rotations). 

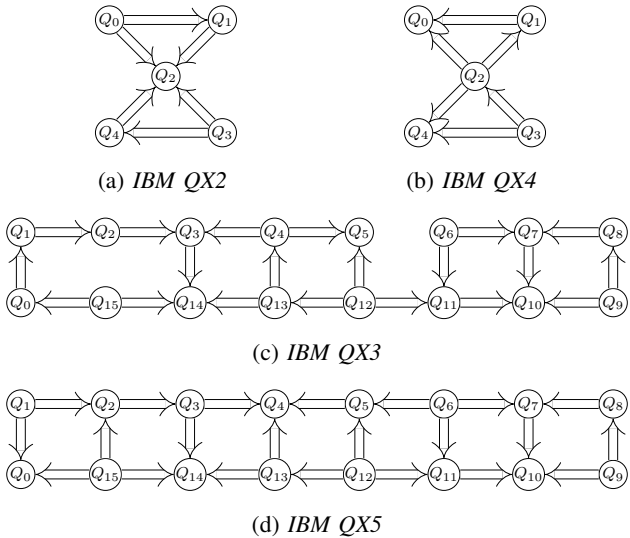
\begin{figure}
	\centering
	\begin{subfigure}[b]{0.45\linewidth}
		\centering
		\scalebox{0.75}{
		\begin{tikzpicture}[terminal/.style={draw,rectangle,inner sep=2pt}]
		\matrix[matrix of nodes,ampersand replacement=\&,every node/.style={vertex},column sep={1cm,between origins},row sep={1.cm,between origins}] (qmdd) {
			\node (n0) {$Q_0$}; \& \& \node (n1) {$Q_1$}; \\
			\& \node (n2) {$Q_2$}; \&  \\
			\node (n4) {$Q_4$}; \& \& \node (n3) {$Q_3$}; \\
		};
		\draw[-Implies,line width=0.5pt,double distance=4pt] (n0.east) -- (n1.west);
		\draw[-Implies,line width=0.5pt,double distance=4pt] (n0.south east) -- (n2.north west);
		\draw[-Implies,line width=0.5pt,double distance=4pt] (n1.south west) -- (n2.north east);
		\draw[-Implies,line width=0.5pt,double distance=4pt] (n3.north west) -- (n2.south east);
		\draw[-Implies,line width=0.5pt,double distance=4pt] (n3.west) -- (n4.east);
		\draw[-Implies,line width=0.5pt,double distance=4pt] (n4.north east) -- (n2.south west);
		\end{tikzpicture}}
			
		\caption{\emph{IBM QX2}}
	\end{subfigure}
	\begin{subfigure}[b]{0.45\linewidth}
	\centering
	\scalebox{0.75}{
		\begin{tikzpicture}[terminal/.style={draw,rectangle,inner sep=2pt}]
		\matrix[matrix of nodes,ampersand replacement=\&,every node/.style={vertex},column sep={1cm,between origins},row sep={1.cm,between origins}] (qmdd) {
			\node (n0) {$Q_0$}; \& \& \node (n1) {$Q_1$}; \\
			\& \node (n2) {$Q_2$}; \&  \\
			\node (n4) {$Q_4$}; \& \& \node (n3) {$Q_3$}; \\
		};
		\draw[-Implies,line width=0.5pt,double distance=4pt] (n1.west) -- (n0.east);
		\draw[-Implies,line width=0.5pt,double distance=4pt] (n2.north west) -- (n0.south east);
		\draw[-Implies,line width=0.5pt,double distance=4pt] (n2.north east) -- (n1.south west);
		\draw[-Implies,line width=0.5pt,double distance=4pt] (n2.south west) -- (n4.north east);
		\draw[-Implies,line width=0.5pt,double distance=4pt] (n3.west) -- (n4.east);
		\draw[-Implies,line width=0.5pt,double distance=4pt] (n3.north west) -- (n2.south east);
		\end{tikzpicture}}
	
	\caption{\emph{IBM QX4}}
\end{subfigure}

\vspace*{2mm}
	\begin{subfigure}[b]{\linewidth}
		\centering
		\scalebox{0.75}{
	\begin{tikzpicture}[terminal/.style={draw,rectangle,inner sep=2pt}]
	\matrix[matrix of nodes,ampersand replacement=\&,every node/.style={vertex},column sep={1.5cm,between origins},row sep={1.25cm,between origins}] (qmdd) {
		\node (n1) {$Q_1$}; \& \node (n2) {$Q_{2}$}; \& \node (n3) {$Q_{3}$}; \& \node (n4) {$Q_{4}$}; \& \node (n5) {$Q_{5}$}; \& \node (n6) {$Q_{6}$}; \& \node (n7) {$Q_{7}$}; \& \node (n8) {$Q_{8}$}; \\
		\node (n0) {$Q_0$}; \& \node (n15) {$Q_{15}$}; \& \node (n14) {$Q_{14}$}; \& \node (n13) {$Q_{13}$}; \& \node (n12) {$Q_{12}$}; \& \node (n11) {$Q_{11}$}; \& \node (n10) {$Q_{10}$}; \& \node (n9) {$Q_{9}$}; \\
		};
		\draw[-Implies,line width=0.5pt,double distance=4pt] (n0.north)--(n1.south);
		\draw[-Implies,line width=0.5pt,double distance=4pt] (n1.east)--(n2.west);
		\draw[-Implies,line width=0.5pt,double distance=4pt] (n2.east)--(n3.west);
		\draw[-Implies,line width=0.5pt,double distance=4pt] (n3.south)--(n14.north);
		\draw[-Implies,line width=0.5pt,double distance=4pt] (n4.west)--(n3.east);
		\draw[-Implies,line width=0.5pt,double distance=4pt] (n4.east)--(n5.west);
		\draw[-Implies,line width=0.5pt,double distance=4pt] (n6.east)--(n7.west);
		\draw[-Implies,line width=0.5pt,double distance=4pt] (n6.south)--(n11.north);
		\draw[-Implies,line width=0.5pt,double distance=4pt] (n7.south)--(n10.north);
		\draw[-Implies,line width=0.5pt,double distance=4pt] (n8.west)--(n7.east);
		\draw[-Implies,line width=0.5pt,double distance=4pt] (n9.north)--(n8.south);
		\draw[-Implies,line width=0.5pt,double distance=4pt] (n9.west)--(n10.east);
		\draw[-Implies,line width=0.5pt,double distance=4pt] (n11.east)--(n10.west);
		\draw[-Implies,line width=0.5pt,double distance=4pt] (n12.east)--(n11.west);
		\draw[-Implies,line width=0.5pt,double distance=4pt] (n12.north)--(n5.south);
		\draw[-Implies,line width=0.5pt,double distance=4pt] (n12.west)--(n13.east);
		\draw[-Implies,line width=0.5pt,double distance=4pt] (n13.north)--(n4.south);
		\draw[-Implies,line width=0.5pt,double distance=4pt] (n13.west)--(n14.east);
		\draw[-Implies,line width=0.5pt,double distance=4pt] (n15.east)--(n14.west);
		\draw[-Implies,line width=0.5pt,double distance=4pt] (n15.west)--(n0.east);
	\end{tikzpicture}
}
	\caption{\emph{IBM QX3}}\label{fig:qx3arch}
\end{subfigure}

\vspace*{2mm}
	\begin{subfigure}[b]{\linewidth}
	\centering
	\scalebox{0.75}{
		\begin{tikzpicture}[terminal/.style={draw,rectangle,inner sep=2pt}]
		\matrix[matrix of nodes,ampersand replacement=\&,every node/.style={vertex},column sep={1.5cm,between origins},row sep={1.25cm,between origins}] (qmdd) {
			\node (n1) {$Q_1$}; \& \node (n2) {$Q_{2}$}; \& \node (n3) {$Q_{3}$}; \& \node (n4) {$Q_{4}$}; \& \node (n5) {$Q_{5}$}; \& \node (n6) {$Q_{6}$}; \& \node (n7) {$Q_{7}$}; \& \node (n8) {$Q_{8}$}; \\
			\node (n0) {$Q_0$}; \& \node (n15) {$Q_{15}$}; \& \node (n14) {$Q_{14}$}; \& \node (n13) {$Q_{13}$}; \& \node (n12) {$Q_{12}$}; \& \node (n11) {$Q_{11}$}; \& \node (n10) {$Q_{10}$}; \& \node (n9) {$Q_{9}$}; \\
		};
		\draw[-Implies,line width=0.5pt,double distance=4pt] (n1.south)--(n0.north);
		\draw[-Implies,line width=0.5pt,double distance=4pt] (n1.east)--(n2.west);
		\draw[-Implies,line width=0.5pt,double distance=4pt] (n2.east)--(n3.west);		
		\draw[-Implies,line width=0.5pt,double distance=4pt] (n3.south)--(n14.north);
		\draw[-Implies,line width=0.5pt,double distance=4pt] (n3.east)--(n4.west);
		\draw[-Implies,line width=0.5pt,double distance=4pt] (n5.west)--(n4.east);
		\draw[-Implies,line width=0.5pt,double distance=4pt] (n6.west)--(n5.east);
		\draw[-Implies,line width=0.5pt,double distance=4pt] (n6.east)--(n7.west);
		\draw[-Implies,line width=0.5pt,double distance=4pt] (n6.south)--(n11.north);
		\draw[-Implies,line width=0.5pt,double distance=4pt] (n7.south)--(n10.north);
		\draw[-Implies,line width=0.5pt,double distance=4pt] (n8.west)--(n7.east);
		\draw[-Implies,line width=0.5pt,double distance=4pt] (n9.north)--(n8.south);
		\draw[-Implies,line width=0.5pt,double distance=4pt] (n9.west)--(n10.east);
		\draw[-Implies,line width=0.5pt,double distance=4pt] (n11.east)--(n10.west);
		\draw[-Implies,line width=0.5pt,double distance=4pt] (n12.east)--(n11.west);
		\draw[-Implies,line width=0.5pt,double distance=4pt] (n12.north)--(n5.south);
		\draw[-Implies,line width=0.5pt,double distance=4pt] (n12.west)--(n13.east);
		\draw[-Implies,line width=0.5pt,double distance=4pt] (n13.north)--(n4.south);
		\draw[-Implies,line width=0.5pt,double distance=4pt] (n13.west)--(n14.east);
		\draw[-Implies,line width=0.5pt,double distance=4pt] (n15.east)--(n14.west);
		\draw[-Implies,line width=0.5pt,double distance=4pt] (n15.west)--(n0.east);
		\draw[-Implies,line width=0.5pt,double distance=4pt] (n15.north)--(n2.south);
		\end{tikzpicture}
	}
	\caption{\emph{IBM QX5}}\label{fig:qx5arch}
\end{subfigure}

	\caption{Coupling map of the \emph{IBM QX} architectures~\cite{qxbackends}}
	\label{fig:coupling}
\end{figure}

However, there are significant restrictions which have to be satisfied when running quantum algorithms on these architectures. 
In fact, the user first has to decompose all \mbox{non-elementary} quantum operations (e.g.~Toffoli gate, SWAP gate, or Fredkin gate) to the elementary operations $U(\theta,\phi,\lambda)$ and $CNOT$. 
Moreover, two-qubit gates, i.e.~CNOT gates, cannot arbitrarily be placed in the architecture but are restricted to dedicated pairs of qubits only.
Even within these pairs, it is firmly defined which qubit is supposed to work as target and which qubit is supposed to work as control. 
These restrictions are given by the so-called \emph{coupling-map} illustrated in Fig.~\ref{fig:coupling}, which sketches the layout of the currently available \emph{IBM QX} architectures. The circles indicate physical qubits (denoted by~$Q_i$) and arrows indicate the possible CNOT applications, i.e.~an 
arrow pointing from physical qubit $Q_i$ to qubit $Q_j$ defines that a CNOT with control qubit $Q_i$ and target qubit $Q_j$ can be applied.
In the following, these restrictions are called \emph{CNOT-constraints} and need to be satisfied in order to execute a quantum circuit on an QX architecture.

\section{Mapping of Quantum Circuits\\to the IBM QX Architectures}\label{sec:mapping}

Mapping quantum circuits to the \emph{IBM~QX} architectures requires the consideration of two major issues.
On the one hand, all gates of the given quantum circuit to be mapped have to be decomposed to elementary operations supported by the hardware, i.e.~CNOTs and parameterized U gates. On the other hand, the $n$ logical qubits $q_0,q_1,\ldots q_{n-1}$ of that quantum circuit have to be mapped to the $m$ physical qubits $Q_0,Q_1,\ldots Q_{m-1}$ ($m=5$ for \emph{QX2} and \emph{QX4}, whereas $m=16$ for \emph{QX3} and \emph{QX5}) of the \emph{IBM QX} architecture. Each logical qubit has to be represented by a physical one, such that all CNOT-constraints are satisfied. 
In this section, we describe how these two issues can be handled in an automatic fashion, what problems occur during this process, and how they can be addressed.

\subsection{Decomposing Quantum Circuits to Elementary Operations}
\label{sec:decomp}

Considering the first issue, IBM has developed the quantum assembly language OpenQASM~\cite{cross2017open} that supports specification of quantum circuits.
Besides elementary gates, the language allows the definition of complex gates that are composed from the elementary operations CNOT and $U$. These gates can then be nested to define even more complex gates. Consequently, as long as a decomposition of the gates used in a description of the desired quantum functionality are provided by the circuit designer, the nested structures are just flattened during the mapping process. 

In case the desired quantum functionality is not provided in OpenQASM, decomposition or synthesis approaches such as those proposed in~\cite{DBLP:journals/tcad/AmyMMR13,MWZ:2011,matsumoto2008representation,WSOD:2013} and~\cite{MMD:2003,DBLP:journals/tcad/ZulehnerW18,DBLP:conf/date/NiemannWD18}, respectively can be applied which determine (e.g.~depth optimal) realizations of quantum functionality for specific libraries like \emph{Clifford+T}~\cite{DBLP:journals/tcad/AmyMMR13} or \emph{NCV}~\cite{BBC+:95}. They typically use search algorithms or a matrix representation of the quantum functionality. For the \emph{Clifford+T} library, Matsumoto and Amano developed a normal form for single qubit operations~\cite{matsumoto2008representation}, which allows for a unique and T-depth optimal decomposition (approximation) of arbitrary single qubit gates (e.g.~rotations) into a sequence of Clifford+T gates (up to a certain error $\epsilon$).
Several such automated methods are available in Quipper (a functional programming language for quantum computing~\cite{DBLP:conf/pldi/GreenLRSV13}), the ScaffCC compiler for the Scaffold language~\cite{abhari2012scaffold,DBLP:conf/cf/JavadiAbhariPKHLCM14}, and RevKit~\cite{SFWD:2010}. 
Since IBM provides the decomposition for commonly used gates like the Clifford+T gates, (controlled) rotations, or Toffoli gates to their gate library, these approaches can be utilized.

\begin{example}
	\label{ex:swap}
	One commonly used operation is the SWAP operation, which exchanges the states of two qubits. Since the SWAP operation is not part of the gate library of IBM's QX architectures, it has to be decomposed into single-qubit gates and CNOTs as shown in Fig.~\ref{fig:swap_decomposition}. Assume that logical qubits $q_0$ and $q_1$ are initially mapped to the physical qubits $Q_0$ and $Q_1$ of QX2, and that their values are to be swapped. As a first decomposition step, we realize the SWAP operation with three CNOTs. 
	If we additionally consider the \mbox{CNOT-constraints}, we have to flip the direction of the CNOT in the middle.
	To this end, we apply Hadamard operations before and after this CNOT. These Hadamard operations then have to be realized by the gate $U(\pi/2, 0, \pi) = H$. 
\end{example}

\begin{figure}
	\centering
	\begin{tikzpicture}
	\node (n1) {
		\begin{tikzpicture}
		\draw[line width=0.300000] (0.500000,0.750000) -- (1.000000,0.750000);
		\draw (0.400000,0.750000) node [left] {$Q_0\leftarrow q_0$};
		\draw (1.100000,0.750000) node [right] {$q_1$};
		\draw[line width=0.300000] (0.500000,0.250000) -- (1.000000,0.250000);
		\draw (0.400000,0.250000) node [left] {$Q_1\leftarrow q_1$};
		\draw (1.10000,0.250000) node [right] {$q_0$};
		
		\draw[line width=0.300000] (0.750000,0.250000) -- (0.750000,0.750000);
		\draw[line width=0.3] (0.65,0.65) -- ++(0.2,0.2) (0.65,0.85) -- ++(0.2,-0.2);
		\draw[line width=0.3] (0.65,0.15) -- ++(0.2,0.2) (0.65,0.35) -- ++(0.2,-0.2);
		
		\end{tikzpicture}
	};
	
	\node[right=1cm of n1] (n2) {
		\begin{tikzpicture}
		\draw[line width=0.300000] (0.500000,0.750000) -- (2.000000,0.750000);
		\draw (0.400000,0.750000) node [left] {$Q_0\leftarrow q_0$};
		\draw (2.100000,0.750000) node [right] {$q_1$};
		\draw[line width=0.300000] (0.500000,0.250000) -- (2.0000,0.250000);
		\draw (0.400000,0.250000) node [left] {$Q_1 \leftarrow q_1$};
		\draw (2.10000,0.250000) node [right] {$q_0$};

		\draw[line width=0.300000] (0.750000,0.75000) -- (0.750000,0.050000);
		\draw[fill] (0.750000,0.750000) circle (0.100000);
		\draw[line width=0.3] (0.75, 0.25) circle (0.2);
		
		\draw[line width=0.300000] (1.250000,0.95000) -- (1.250000,0.250000);
		\draw[fill] (1.250000,0.250000) circle (0.100000);
		\draw[line width=0.3] (1.25, 0.75) circle (0.2);
		
		\draw[line width=0.300000] (1.750000,0.75000) -- (1.750000,0.050000);
		\draw[fill] (1.750000,0.750000) circle (0.100000);
		\draw[line width=0.3] (1.75, 0.25) circle (0.2);
		
		\end{tikzpicture}
	};

	\node[anchor = north,yshift=-0.75cm] at ($(n1.west)!.5!(n2.east)$) (n3) {
		\begin{tikzpicture}
		\draw[line width=0.300000] (0.500000,0.750000) -- (3.000000,0.750000);
		\draw (0.400000,0.750000) node [left] {$Q_0\leftarrow q_0$};
		\draw (3.100000,0.750000) node [right] {$q_1$};
		\draw[line width=0.300000] (0.500000,0.250000) -- (3.0000,0.250000);
		\draw (0.400000,0.250000) node [left] {$Q_1\leftarrow q_1$};
		\draw (3.10000,0.250000) node [right] {$q_0$};

		\draw[line width=0.300000] (0.750000,0.75000) -- (0.750000,0.050000);
		\draw[fill] (0.750000,0.750000) circle (0.100000);
		\draw[line width=0.3] (0.75, 0.25) circle (0.2);
		
		\draw[line width=0.3,fill=white] (1.05,0.55) rectangle ++(0.4,0.4); \node at (1.25,0.95) {\sf H};
		\draw[line width=0.3,fill=white] (1.05,0.05) rectangle ++(0.4,0.4); \node at (1.25,0.45) {\sf H};
		
		\draw[line width=0.300000] (1.750000,0.75000) -- (1.750000,0.050000);
		\draw[fill] (1.750000,0.750000) circle (0.100000);
		\draw[line width=0.3] (1.75, 0.25) circle (0.2);
		
		\draw[line width=0.3,fill=white] (2.05,0.55) rectangle ++(0.4,0.4); \node at (2.25,0.95) {\sf H};
		\draw[line width=0.3,fill=white] (2.05,0.05) rectangle ++(0.4,0.4); \node at (2.25,0.45) {\sf H};
		
		\draw[line width=0.300000] (2.750000,0.75000) -- (2.750000,0.050000);
		\draw[fill] (2.750000,0.750000) circle (0.100000);
		\draw[line width=0.3] (2.75, 0.25) circle (0.2);
		
		\end{tikzpicture}
	};
	
\end{tikzpicture}
\caption{Decomposition of SWAP gates}
\label{fig:swap_decomposition}
\end{figure}
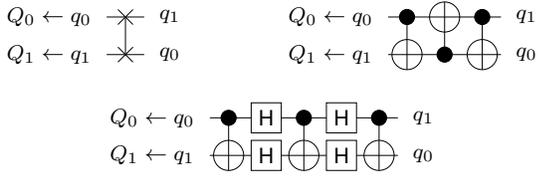

\begin{figure*}
	
	\begin{subfigure}[b]{0.2\textwidth}
		\centering
		\scalebox{0.9}{
			\begin{tikzpicture}
			
			\draw[line width=0.300000] (0.500000,2.750000) -- (3.000000,2.750000);
			\draw (0.400000,2.750000) node [left] {$q_0$};
			\draw (3.100000,2.750000) node [right] {$q_0$};
			\draw[line width=0.300000] (0.500000,2.250000) -- (3.000000,2.250000);
			\draw (0.400000,2.250000) node [left] {$q_1$};
			\draw (3.100000,2.250000) node [right] {$q_1$};
			\draw[line width=0.300000] (0.500000,1.750000) -- (3.000000,1.750000);
			\draw (0.400000,1.750000) node [left] {$q_2$};
			\draw (3.100000,1.750000) node [right] {$q_2$};
			\draw[line width=0.300000] (0.500000,1.250000) -- (3.000000,1.250000);
			\draw (0.400000,1.250000) node [left] {$q_3$};
			\draw (3.100000,1.250000) node [right] {$q_3$};
			\draw[line width=0.300000] (0.500000,0.750000) -- (3.000000,0.750000);
			\draw (0.400000,0.750000) node [left] {$q_4$};
			\draw (3.100000,0.750000) node [right] {$q_4$};
			\draw[line width=0.300000] (0.500000,0.250000) -- (3.000000,0.250000);
			\draw (0.400000,0.250000) node [left] {$q_5$};
			\draw (3.100000,0.250000) node [right] {$q_5$};
						
			\draw[line width=0.300000] (0.7500,1.05000) -- (0.75,1.750000);
			\draw[fill] (0.75,1.750000) circle (0.100000);
			\draw[line width=0.3] (0.75, 1.25) circle (0.2);
			
			\draw[line width=0.300000] (1.2500,2.25000) -- (1.25000,2.950000);
			\draw[fill] (1.2500,2.250000) circle (0.100000);
			\draw[line width=0.3] (1.25, 2.75) circle (0.2);
			
			\draw[line width=0.300000] (1.7500,0.55000) -- (1.75000,2.250000);
			\draw[fill] (1.7500,2.250000) circle (0.100000);
			\draw[line width=0.3] (1.75, 0.75) circle (0.2);
			
			\draw[line width=0.300000] (2.2500,0.25000) -- (2.25000,1.450000);
			\draw[fill] (2.2500,0.250000) circle (0.100000);
			\draw[line width=0.3] (2.25, 1.25) circle (0.2);
			
			\draw[line width=0.300000] (2.75,1.75000) -- (2.75,1.050000);
			\draw[fill] (2.75,1.750000) circle (0.100000);
			\draw[line width=0.3] (2.75, 1.25) circle (0.2);
			
			\node at (0.75,0) {$g_0$};
			\node at (1.25,0) {$g_1$};
			\node at (1.75,0) {$g_2$};
			\node at (2.25,0) {$g_3$};
			\node at (2.75,0) {$g_4$};
			
			\draw[dashed] (1.5,-0.25) -- (1.5,3.25);
			\draw[dashed] (2.5,-0.25) -- (2.5,3.25);
			
			\node at (1,3.1) {$l_0$};
			\node at (2,3.1) {$l_1$};
			\node at (2.75,3.1) {$l_2$};
			
			\end{tikzpicture}}
		\caption{Original circuit}
		\label{fig:circuit1}
	\end{subfigure}\hfil
	\begin{subfigure}[b]{0.45\textwidth}
		\centering
		\scalebox{0.9}{
			\begin{tikzpicture}
			
			\draw[white] (1.5,-0.25) -- (1.5,3.25);

			\draw[line width=0.300000] (0.500000,2.750000) -- (6.5000000,2.750000);
			\draw (0.400000,2.750000) node [left] {$Q_0\leftarrow q_0$};
			\draw (6.600000,2.750000) node [right] {$q_0$};
			\draw[line width=0.300000] (0.500000,2.250000) -- (6.5000000,2.250000);
			\draw (0.400000,2.250000) node [left] {$Q_1\leftarrow q_1$};
			\draw (6.600000,2.250000) node [right] {$q_2$};
			\draw[line width=0.300000] (0.500000,1.750000) -- (6.5000000,1.750000);
			\draw (0.400000,1.750000) node [left] {$Q_2\leftarrow q_2$};
			\draw (6.600000,1.750000) node [right] {$q_3$};
			\draw[line width=0.300000] (0.500000,1.250000) -- (6.5000000,1.250000);
			\draw (0.400000,1.250000) node [left] {$Q_3\leftarrow q_3$};
			\draw (6.600000,1.250000) node [right] {$q_1$};
			\draw[line width=0.300000] (0.500000,0.750000) -- (6.5000000,0.750000);
			\draw (0.400000,0.750000) node [left] {$Q_{14}\leftarrow q_4$};
			\draw (6.600000,0.750000) node [right] {$q_4$};
			\draw[line width=0.300000] (0.500000,0.250000) -- (6.5000000,0.250000);
			\draw (0.400000,0.250000) node [left] {$Q_{15}\leftarrow q_5$};
			\draw (6.600000,0.250000) node [right] {$q_5$};
			
			\draw[line width=0.3,fill=white] (0.55,2.55) rectangle ++(0.4,0.4); \node at (0.75,2.75) {\sf H};
			\draw[line width=0.3,fill=white] (0.55,2.05) rectangle ++(0.4,0.4); \node at (0.75,2.25) {\sf H};
			
			\draw[line width=0.300000] (0.75,1.05000) -- (0.75,1.750000);
			\draw[fill] (0.75,1.750000) circle (0.100000);
			\draw[line width=0.3] (0.75, 1.25) circle (0.2);
			
			\draw[line width=0.300000] (1.2500,2.05000) -- (1.25000,2.750000);
			\draw[fill] (1.2500,2.750000) circle (0.100000);
			\draw[line width=0.3] (1.25, 2.25) circle (0.2);
			
			\draw[line width=0.3,fill=white] (1.55,2.55) rectangle ++(0.4,0.4); \node at (1.75,2.75) {\sf H};
			\draw[line width=0.3,fill=white] (1.55,2.05) rectangle ++(0.4,0.4); \node at (1.75,2.25) {\sf H};
			
			\draw[line width=0.300000] (2.250000,1.750000) -- (2.250000,2.250000);
			\draw[line width=0.3] (2.15,2.15) -- ++(0.2,0.2) node[xshift=0.1cm, yshift=0.05cm] () {\tiny$q_2$} (2.15,2.35) -- ++(0.2,-0.2);
			\draw[line width=0.3] (2.15,1.65) -- ++(0.2,0.2) node[xshift=0.1cm, yshift=0.05cm] () {\tiny$q_1$} (2.15,1.85) -- ++(0.2,-0.2);
			
			\draw[line width=0.300000] (2.750000,1.250000) -- (2.750000,1.750000);
			\draw[line width=0.3] (2.65,1.65) -- ++(0.2,0.2) node[xshift=0.1cm, yshift=0.05cm] () {\tiny$q_3$} (2.65,1.85) -- ++(0.2,-0.2);
			\draw[line width=0.3] (2.65,1.15) -- ++(0.2,0.2) node[xshift=0.1cm, yshift=0.05cm] () {\tiny$q_1$} (2.65,1.35) -- ++(0.2,-0.2);
			
			\draw[line width=0.300000] (3.2500,0.55000) -- (3.25000,1.250000);
			\draw[fill] (3.2500,1.250000) circle (0.100000);
			\draw[line width=0.3] (3.25, 0.75) circle (0.2);
			
			\draw[line width=0.300000] (3.750000,1.250000) -- (3.750000,1.750000);
			\draw[line width=0.3] (3.65,1.65) -- ++(0.2,0.2) node[xshift=0.1cm, yshift=0.05cm] () {\tiny$q_1$} (3.65,1.85) -- ++(0.2,-0.2);
			\draw[line width=0.3] (3.65,1.15) -- ++(0.2,0.2) node[xshift=0.1cm, yshift=0.05cm] () {\tiny$q_3$} (3.65,1.35) -- ++(0.2,-0.2);
			
			\draw[line width=0.300000] (4.250000,1.250000) -- (4.250000,0.750000);
			\draw[line width=0.3] (4.15,1.15) -- ++(0.2,0.2) node[xshift=0.1cm, yshift=0.05cm] () {\tiny$q_4$} (4.15,1.35) -- ++(0.2,-0.2);
			\draw[line width=0.3] (4.15,0.65) -- ++(0.2,0.2) node[xshift=0.1cm, yshift=0.05cm] () {\tiny$q_3$} (4.15,0.85) -- ++(0.2,-0.2);
			
			\draw[line width=0.300000] (4.7500,0.25000) -- (4.75000,0.950000);
			\draw[fill] (4.7500,0.250000) circle (0.100000);
			\draw[line width=0.3] (4.75, 0.75) circle (0.2);
			
			\draw[line width=0.300000] (5.250000,1.250000) -- (5.250000,0.750000);
			\draw[line width=0.3] (5.15,1.15) -- ++(0.2,0.2) node[xshift=0.1cm, yshift=0.05cm] () {\tiny$q_3$} (5.15,1.35) -- ++(0.2,-0.2);
			\draw[line width=0.3] (5.15,0.65) -- ++(0.2,0.2) node[xshift=0.1cm, yshift=0.05cm] () {\tiny$q_4$} (5.15,0.85) -- ++(0.2,-0.2);
			
			\draw[line width=0.300000] (5.750000,1.250000) -- (5.750000,1.750000);
			\draw[line width=0.3] (5.65,1.65) -- ++(0.2,0.2) node[xshift=0.1cm, yshift=0.05cm] () {\tiny$q_3$} (5.65,1.85) -- ++(0.2,-0.2);
			\draw[line width=0.3] (5.65,1.15) -- ++(0.2,0.2) node[xshift=0.1cm, yshift=0.05cm] () {\tiny$q_1$} (5.65,1.35) -- ++(0.2,-0.2);
			
			\draw[line width=0.300000] (6.25,2.25000) -- (6.25,1.550000);
			\draw[fill] (6.25,2.250000) circle (0.100000);
			\draw[line width=0.3] (6.25, 1.75) circle (0.2);
			
			\node at (0.75,0) {$g_0$};
			\node at (1.25,0) {$g_1$};
			\node at (3.25,0) {$g_2$};
			\node at (4.75,0) {$g_3$};
			\node at (6.25,0) {$g_4$};

			\end{tikzpicture}
		}
		\caption{Naive strategy}
		\label{fig:circuit2}
	\end{subfigure}\hfil
	\begin{subfigure}[b]{0.3\textwidth}
		\centering
		\scalebox{0.9}{
			\begin{tikzpicture}
			\draw[white] (1.5,-0.25) -- (1.5,3.25);

			\draw[line width=0.300000] (0.500000,2.750000) -- (4,2.750000);
			\draw (0.400000,2.750000) node [left] {$Q_0\leftarrow q_2$};
			\draw (4.10000,2.750000) node [right] {$q_3$};
			\draw[line width=0.300000] (0.500000,2.250000) -- (4.000000,2.250000);
			\draw (0.400000,2.250000) node [left] {$Q_1\leftarrow q_3$};
			\draw (4.100000,2.250000) node [right] {$q_2$};
			\draw[line width=0.300000] (0.500000,1.750000) -- (4.000000,1.750000);
			\draw (0.400000,1.750000) node [left] {$Q_2\leftarrow q_1$};
			\draw (4.100000,1.750000) node [right] {$q_1$};
			\draw[line width=0.300000] (0.500000,1.250000) -- (4.000000,1.250000);
			\draw (0.400000,1.250000) node [left] {$Q_3\leftarrow q_0$};
			\draw (4.100000,1.250000) node [right] {$q_4$};
			\draw[line width=0.300000] (0.500000,0.750000) -- (4.000000,0.750000);
			\draw (0.400000,0.750000) node [left] {$Q_{4}\leftarrow q_4$};
			\draw (4.100000,0.750000) node [right] {$q_0$};
			\draw[line width=0.300000] (0.500000,0.250000) -- (4.000000,0.250000);
			\draw (0.400000,0.250000) node [left] {$Q_{15}\leftarrow q_5$};
			\draw (4.100000,0.250000) node [right] {$q_5$};
						
			\draw[line width=0.300000] (0.75,1.05000) -- (0.75,1.750000);
			\draw[fill] (0.75,1.750000) circle (0.100000);
			\draw[line width=0.3] (0.75, 1.25) circle (0.2);
			
			\draw[line width=0.300000] (0.7500,2.05000) -- (0.75000,2.750000);
			\draw[fill] (0.7500,2.750000) circle (0.100000);
			\draw[line width=0.3] (0.75, 2.25) circle (0.2);
			
			\draw[line width=0.300000] (1.250000,2.250000) -- (1.250000,2.750000);
			\draw[line width=0.3] (1.15,2.65) -- ++(0.2,0.2) node[xshift=0.1cm, yshift=0.05cm] () {\tiny$q_3$} (1.15,2.85) -- ++(0.2,-0.2);
			\draw[line width=0.3] (1.15,2.15) -- ++(0.2,0.2) node[xshift=0.1cm, yshift=0.05cm] () {\tiny$q_2$} (1.15,2.35) -- ++(0.2,-0.2);
			
			\draw[line width=0.300000] (1.250000,1.250000) -- (1.250000,0.750000);
			\draw[line width=0.3] (1.15,1.15) -- ++(0.2,0.2) node[xshift=0.1cm, yshift=0.05cm] () {\tiny$q_4$} (1.15,1.35) -- ++(0.2,-0.2);
			\draw[line width=0.3] (1.15,0.65) -- ++(0.2,0.2) node[xshift=0.1cm, yshift=0.05cm] () {\tiny$q_0$} (1.15,0.85) -- ++(0.2,-0.2);
			
			\draw[line width=0.300000] (1.7500,1.05000) -- (1.75000,1.750000);
			\draw[fill] (1.7500,1.750000) circle (0.100000);
			\draw[line width=0.3] (1.75, 1.25) circle (0.2);
			
			\draw[line width=0.300000] (2.2500,0.25000) -- (2.25000,2.950000);
			\draw[fill] (2.2500,0.250000) circle (0.100000);
			\draw[line width=0.3] (2.25, 2.75) circle (0.2);
			
			\draw[line width=0.3,fill=white] (2.55,2.55) rectangle ++(0.4,0.4); \node at (2.75,2.75) {\sf H};
			\draw[line width=0.3,fill=white] (2.55,2.05) rectangle ++(0.4,0.4); \node at (2.75,2.25) {\sf H};
			
			\draw[line width=0.300000] (3.2500,2.05000) -- (3.25000,2.750000);
			\draw[fill] (3.2500,2.750000) circle (0.100000);
			\draw[line width=0.3] (3.25, 2.25) circle (0.2);
			
			\draw[line width=0.3,fill=white] (3.55,2.55) rectangle ++(0.4,0.4); \node at (3.75,2.75) {\sf H};
			\draw[line width=0.3,fill=white] (3.55,2.05) rectangle ++(0.4,0.4); \node at (3.75,2.25) {\sf H};
			
			\node at (0.75,0) {$g_0$,$g_1$};
			\node at (1.75,0) {$g_2$};
			\node at (2.25,0) {$g_3$};
			\node at (3.25,0) {$g_4$};
			
			\end{tikzpicture}}
		\caption{Proposed strategy}
		\label{fig:circuit3}
	\end{subfigure}
	\caption{Mapping of a quantum circuit to the \emph{IBM QX3} architecture}
	\vspace*{-2mm}
\end{figure*}
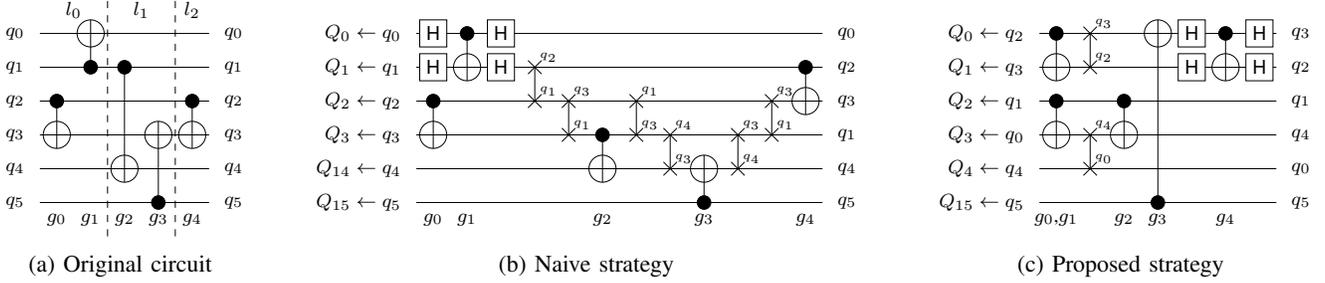

Hence, decomposing the desired quantum functionality to the elementary gate library is already well covered by corresponding related work. 
Unfortunately, this is not the case for the second issue, which is discussed next.

\subsection{Satisfying CNOT-constraints}
\label{sec:cnot-constraints}

Recall that, in order to satisfy the CNOT-constraints as defined in Section~\ref{sec:architecture}, the $n$ logical qubits $q_0,q_1,\ldots q_{n-1}$ of the quantum circuit to be realized have to be mapped to the $m$ physical qubits $Q_0,Q_1,\ldots Q_{m-1}$ ($m=5$ for \emph{QX2} and \emph{QX4}, whereas $m=16$ for \emph{QX3} and \emph{QX5}) of the \mbox{\emph{IBM QX}} architecture. Usually, there exists no mapping solution that satisfies all CNOT-constrains throughout the whole circuit (this is already impossible if CNOT gates are applied to qubit pairs $(q_h,q_i)$, $(q_h,q_j)$, $(q_h,q_k)$, and $(q_h, q_l)$ with $h \neq i\neq j \neq k \neq l$).
That is, whatever initial mapping might be imposed at the beginning, it may have to be changed during the execution of a quantum circuit (namely exactly when a gate is to be executed which violates a CNOT-constraint). To this end, $H$ and SWAP gates can be applied to change the direction of a CNOT gate and to change the mapping of the logical qubits, respectively. 
In other words, these gates can be used to ``move'' around the logical qubits on the actual hardware until the \mbox{CNOT-constraints} are satisfied. An example illustrates the idea.

\begin{example}
	\label{ex:naive}
	Consider the quantum circuit composed of 5 CNOT gates shown in Fig.~\ref{fig:circuit1} and assume that the logical qubits $q_0$, $q_1$, $q_2$, $q_3$, $q_4$, and $q_5$ are respectively mapped to the physical qubits $Q_0$, $Q_1$, $Q_2$, $Q_3$, $Q_{14}$, and $Q_{15}$ of the \emph{IBM QX3} architecture shown in Fig.~\ref{fig:qx3arch}. The first gate can directly be applied, because the CNOT-constraint is satisfied. For the second gate, the direction has to be changed  because a CNOT with control qubit $Q_0$ and target $Q_1$ is valid, but not vice versa. This can be accomplished by inserting Hadamard gates as shown in Fig.~\ref{fig:circuit2}.
	For the third gate, we have to change the mapping. To this end, we insert SWAP operations $SWAP(Q_1, Q_2)$ and $SWAP(Q_2, Q_3)$ to move logical qubit $q_1$ towards logical qubit $q_4$ (see Fig.~\ref{fig:circuit2}).
	Afterwards, $q_1$ and $q_4$ are mapped to the physical qubits $Q_3$ and $Q_{14}$, respectively, which allows us to apply the desired CNOT gate. Following this procedure for the remaining qubits eventually results in the circuit shown in Fig.~\ref{fig:circuit2}.
\end{example}

However, inserting the additional gates in order to satisfy the CNOT-constraints 
drastically increases the number of operations -- a significant drawback which affects the reliability of the quantum circuit since each gate has a certain 
error rate. Since each SWAP operation is composed of 7 elementary gates (cf.~Fig.~\ref{fig:swap_decomposition}), particularly their number shall be kept as small as possible. Besides that, the circuit depth shall be kept as small as it is related to the time required to execute the quantum circuit. 
Since a SWAP operation has a depth of 5, this also motivates the search for alternative solutions which realize a CNOT-constraint-compliant mapping with as few SWAP operations as possible. 

\begin{example}
	Consider again the given quantum circuit from Fig.~\ref{fig:circuit1} as well as its mapping derived in Example~\ref{ex:naive} and shown in Fig.~\ref{fig:circuit2}.
	This circuit	
	is composed of 51 elementary operations and has a depth of 36. 
	In contrast, the same quantum circuit can be realized with only 23 elementary operations and depth of 10 as shown in Fig.~\ref{fig:circuit3} ($g_2$ and $g_3$ can be applied concurrently) -- a significant reduction. 
\end{example}

Determining proper mappings has similarities with recent work on nearest neighbor optimization of quantum circuits proposed in~\cite{DBLP:journals/tcad/WilleLD14,DBLP:journals/qip/SaeediWD11,DBLP:conf/aspdac/WilleKWRCD16,SSP:2013,DBLP:conf/aspdac/ShafaeiSP14,DBLP:conf/rc/WilleQIYM16,zulehner2017exact}.\footnote{These approaches utilize satisfiability solvers, search algorithms, or dedicated data structures to tackle the underlying complexity.} In that work, SWAP gates have also been applied to move qubits together in order to satisfy a physical constraint. However, these works consider simpler and artificial architectures with \mbox{1-dimensional} or \mbox{2-dimensional} layouts where any two-qubit gate can be applied to adjacent qubits.
The CNOT-constraints to be satisfied for the \emph{IBM QX} architectures are much stricter with respect to what physical qubits may interact with each other and also what physical qubit may act as control and as target qubit. Furthermore, the parallel execution of gates (which is possible in the QX architectures) is not considered by these approaches. Besides that, there exists a recent approach that utilizes temporal planning to compile quantum circuits to real architectures~\cite{DBLP:journals/corr/VenturelliDRF17}. However, this approach is rather specialized to \emph{Quantum Alternating Operator Ansatz} (QAOA~\cite{farhi2014quantum}) circuits for solving the MaxCut problem and target the architectures proposed by Rigetti (cf.~\cite{DBLP:conf/icrc/SeteZR16}).
As a consequence, none of the approaches discussed above is directly applicable for the problem considered here.

As a further alternative, IBM provides a solution within its SDK~\cite{qiskit}. This algorithm randomly searches (guided by heuristics) for mappings of the qubits at a certain point of time. These mappings are then realized by adding SWAP gates to the circuit. But this random search is hardly feasible for many quantum circuits and, hence, is not as efficient as circuit designers, e.g.~in the conventional domain, take for granted today. In fact, in many cases the provided method is not capable of determining a CNOT-constraint-compliant mapping within 1 hour (cf.~Section~\ref{sec:results}) -- an issue which will become more serious when further architectures with more qubits are introduced. 

\smallskip

Overall, automatically and efficiently mapping quantum circuits to the \emph{IBM QX} architectures particularily boils down to the question how to efficiently determine a mapping of logical qubits to physical qubits which satisfy the \mbox{CNOT-constraints}. How this problem can be addressed is covered in the next section. 

\section{Efficiently Satisfying CNOT-constraints}
\label{sec:proposed}

In this section, we propose an efficient method for mapping a given quantum circuit (which has already been decomposed into a sequence of elementary gates as described in Section~\ref{sec:decomp}) to the \emph{IBM QX} architectures.
The main objective is to minimize the number of elementary gates which are added in order to make the mapping \mbox{CNOT-constraint-compliant}. Two main steps are employed: First, the given circuit is partitioned into layers which can be realized in a \mbox{CNOT-constraint-compliant} fashion. Afterwards, for each of these layers, a particular compliant mapping is determined which requires as few additional gates as possible. In the following subsections, both steps are described in detail. Afterwards, further optimizations are proposed to reduce the costs of the resulting circuit. 

\subsection{Partitioning the Circuit Into Layers}
\label{sec:depth_part}

As mentioned above, the mapping from logical qubits to physical ones may change over time in order to satisfy all CNOT-constraints, i.e.~the mapping may have to change before a CNOT can be applied. Since each change of the mapping requires additional SWAP operations, we aim for conducting these changes as rarely as possible. 
To this end, we combine gates that can be applied concurrently into so-called \emph{layers} (i.e.~sets of gates). A layer~$l_i$ contains only gates that act on distinct sets of qubits. Furthermore, this allows us to determine a mapping such that the CNOT-constraints for all gates \mbox{$g_j\in l_i$} are satisfied at the same time. We form the layers in a greedy fashion, i.e.~we add a gate to the layer $l_i$ where~$i$ is as small as possible. 
In the circuit diagram representation, this means to move all gates to the left as far as possible without changing the order of gates that share a common qubit. Note that the depth of a circuit 
is equal to the number of layers of a circuit.

\begin{example}
	Consider again the quantum circuit shown in Fig.~\ref{fig:circuit1}.
	The gates of the circuit can be partitioned into three layers  $l_0=\{g_0, g_1\}$, $l_1=\{g_2, g_3\}$, and $l_2=\{g_4\}$ (indicated by the dashed lines in Fig.~\ref{fig:circuit1}).
\end{example}

To satisfy all CNOT constraints, we have to map the logical qubits of each layer $l_i$ to physical ones. 
Since the resulting mapping for layer $l_i$ does not necessarily have to be equal to the mapping determined for the previous layer~$l_{i-1}$, we additionally need to insert SWAP operations that permute the logical qubits from the mapping for layer $l_{i-1}$ to the desired mapping for layer $l_i$. 
In the following, we call this sequence of SWAP operations \emph{permutation layer} $\pi_i$. 
The mapped circuit is then an interleaved sequence of the layers $l_i$ of the original circuit, and the according permutation layers~$\pi_i$, i.e.~$l_0\pi_1l_1\pi_2l_2\ldots$.

\subsection{Determining Compliant Mappings for the Layers}
\label{sec:locally_opt}

For each layer $l_i$, we now determine all mappings \mbox{$\sigma^i_j: \{q_0, q_1, \ldots q_{n-1}\} \rightarrow \{Q_0, Q_1,\ldots Q_{m-1}\}$} 
describing to which physical qubit a logical qubit is mapped. 
The starting point is an initial mapping which is denoted by~$\sigma^i_0$ and obtained from the previous layer $l_{i-1}$, i.e.~$\sigma^i_0 = \hat{\sigma}^{i-1}$ (for~$l_0$, a randomly generated initial mapping that satisfies all CNOT constraints for the gates $g\in l_0$ is used).
Now, this initial mapping~$\sigma^i_0$ should be changed to the desired mapping which is denoted by~$\hat{\sigma}^i$, is CNOT-constraint-compliant for all gates~$g \in l_i$, and can be established from~$\sigma^i_0$ with minimum costs, i.e.~the minimum number of additionally required elementary operations.
In the worst case, determining~$\hat{\sigma}^i$ requires the consideration of  $m!/(m-n)!$ possibilities (where $m$ and $n$ are the number of physical qubits and logical qubits, respectively) -- an exponential complexity.  
We cope with this complexity by applying an $A^*$ search algorithm.

The $A^*$ algorithm~\cite{DBLP:journals/tssc/HartNR68} is a state-space search algorithm. To this end, \mbox{(sub-)solutions} of the considered problem are represented by state nodes. Nodes that represent a solution are called \emph{goal nodes} (multiple goal nodes may exist).
The main idea is to determine the cheapest path (i.e.~the path with the lowest cost) from the root node to a goal node. 
Since the search space is typically exponential, sophisticated mechanisms are employed in order to keep considering as few paths as possible.

All state-space search algorithms are similar in the way they start with a root node (representing an initial partial solution)
which is iteratively expanded towards the goal node (i.e.~the desired complete solution). 
How to choose the node to be expanded next depends on the actual search algorithm. 
For $A^*$ search, we determine the cost of each leaf-node of the search space. Then, the node with the lowest cost is chosen to be expanded next.
To this end, we determine the cost \mbox{$f(x) = g(x) + h(x)$} of a node $x$. The first part ($g(x)$) describes the cost of the current sub-solution (i.e.~the cost of the path from the root to $x$). The second part describes the remaining cost (i.e.~the cost from $x$ to a goal node), which is estimated by a heuristic function $h(x)$. Since the node with the lowest cost is expanded, some parts of the search space (those that lead to expensive solutions) are never expanded.

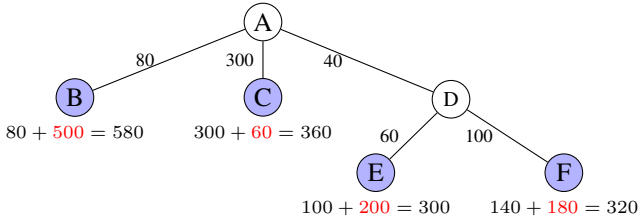
\begin{figure}
	\centering
	\begin{tikzpicture}[terminal/.style={draw,rectangle,inner sep=2pt}]
	\matrix[matrix of nodes,ampersand replacement=\&,every node/.style={vertex},column sep={2.5cm,between origins},row sep={1.0cm,between origins}] (qmdd) {
		\& \node (n1) {\normalsize A}; \& \\
		\node[fill=blue!30] (n2a) {\normalsize B}; \& \node[fill=blue!30] (n2b) {\normalsize C}; \& \node (n2c) {D}; \\
		\& \node[xshift=1.5cm, fill=blue!30] (n3a) {\normalsize E}; \& \node[xshift=1.5cm, fill=blue!30] (n3b) {\normalsize F}; \\			
	};
	
	\draw (n1) -- node[midway, left, xshift=-0.2em] {\scriptsize 80} (n2a);
	\draw (n1) -- node[midway, left, xshift=-0.0em] {\scriptsize 300} (n2b);
	\draw (n1) -- node[midway, left, xshift=-0.2em] {\scriptsize 40} (n2c);
	\draw (n2c) -- node[midway, left, xshift=-0.2em] {\scriptsize 60} (n3a);
	\draw (n2c) -- node[midway, left, xshift=-0.2em] {\scriptsize 100} (n3b);
	
	\node[below = 0cm of n2a.south] (l1) {\scriptsize$80 + \textcolor{red}{500} = 580$};
	\node[below = 0cm of n2b.south] (l2) {\scriptsize$300 + \textcolor{red}{60} = 360$};
	\node[below = 0cm of n3a.south] (l3) {\scriptsize$100 + \textcolor{red}{200} = 300$};
	\node[below = 0cm of n3b.south] (l4) {\scriptsize$140 + \textcolor{red}{180} = 320$};

	\end{tikzpicture}
	
	\caption{$A^*$ search algorithm}
	\label{fig:a*}
\end{figure}

\begin{example}
	Consider the tree shown in Fig.~\ref{fig:a*}. This tree represents the part of the search space that has already been explored for a certain search problem. The nodes that are candidates to be expanded in the next iteration of the $A^*$ algorithm are highlighted in blue. For all these nodes, we determine the cost $f(x) = g(x) + h(x)$. This sum is composed by the cost of the path from the root to the node $x$ (i.e.~the sum of the cost annotated at the respective edges) and the estimated cost of the path from node $x$ to a goal node (provided in red). Consider the node labeled~$E$. This node has cost $f(E) = (40+60) + 200 = 300$. The other candidates labeled $B$, $C$, and $F$ have cost $f(B)=580$, $f(C)=360$, and $f(F)=320$, respectively. Since the node labeled $E$ has the fewest expected cost, it is expanded next. 
\end{example}

Obviously, the heuristic cost should be as accurate as possible, to expand as few nodes as possible. If $h(x)$ always provides the correct minimal remaining cost, only the nodes along the cheapest path from the root node to a goal node would be expanded. But since the minimal costs are usually not known (otherwise, the search problem would be trivial to solve), estimations are employed.
However, to ensure an optimal solution, $h(x)$ has to be \emph{admissible}, i.e.~$h(x)$ must not overestimate the cost of the cheapest path from $x$ to a goal node. This ensures that no goal node is expanded (which terminates the search algorithm) until all nodes that have the potential to lead to a cheaper solution are expanded.

\addtocounter{example}{-1}
\begin{example}[continued]
	Consider again the node labeled $E$. If $h(x)$ is admissible, the true cost of each path from this node to a goal node is greater than or equal to 200.
\end{example}

To use the $A^*$ algorithm for our search problem, an expansion strategy for a state (i.e.~a mapping $\sigma^i_j$) as well as an admissible heuristic function $h(x)$ to estimate the distance of a state to a goal state (i.e.~the mapping $\hat{\sigma}^i$) are required.
Given a mapping $\sigma^i_j$, we can determine all possible successor mappings~$\sigma^i_h$ by employing all possible combinations of SWAP gates that can be applied concurrently.\footnote{Note that we apply multiple SWAP gates concurrently in order to minimize the circuit depth as second criterion (if two solutions require the same number of additional operations).} The fixed costs of all these successor states~$\sigma^i_h$ is then \mbox{$f(\sigma^i_h) = f(\sigma^i_j) + 7 \cdot \#SWAPS$} since each SWAP gate is composed of 7 elementary operations (3 CNOTs and 4 Hadamard operations).
Note that we can restrict the expansion strategy to SWAP operations that affect at least one qubit that occurs in a CNOT gate $g \in l_i$ on layer $l_i$. This is justified by the fact that only these qubits influence whether or not the resulting successor mapping is CNOT-constraint-compliant.

\begin{example}
	\label{ex:locally_opt}
	Consider again the quantum circuit shown in Fig.~\ref{fig:circuit1} and assume we are searching for a mapping for layer~$l_1 = \{g_2, g_3\}$. In the previous layer~$l_0$, the logical qubits~$q_1$, $q_3$, $q_4$, and $q_5$ have been mapped to the physical qubits $Q_0$, $Q_3$, $Q_{14}$, and $Q_{15}$, respectively (i.e.~$\hat{\sigma}^{0}$). This initial mapping $\sigma^1_0 = \hat{\sigma}^0$ does not satisfy the CNOT-constraints for the gates in $l_1$. Since we only consider four qubits in the CNOTs of $l_1$, $\sigma^i_0$ has only 51 successors $\sigma^i_j$.
\end{example}

As mentioned above, to obtain an optimal mapping (i.e.~the mapping with the fewest additionally required elementary operations that satisfies all CNOT-constraints), we need a heuristic function that does not overestimate the real cost (i.e.~the minimum number of additionally inserted elementary operations) for reaching $\hat{\sigma}^i$ from $\sigma^i_j$. 

The real minimum costs for an individual CNOT gate~$g \in l_i$ can easily be  determined given $\sigma^i_j$.
First, we determine the physical qubits $Q_s$ and $Q_t$ to which the control and target qubit of $g$ are mapped (which is given by $\sigma^i_j$). 
Using the coupling map of the architecture (cf.~Fig.~\ref{fig:coupling}), we then determine the shortest path (following the arrows in the coupling map\footnote{The direction of the arrow does not matter since a SWAP can be applied beween two physical qubits iff a CNOT can be applied.}) $\hat{p}$ from $Q_s$ to $Q_t$. 
The costs of the CNOT gate~\mbox{$h(g, \sigma^i_j) = (\left|\hat{p}\right|-1) \cdot 7$} are then determined by the length of this shortest path $\left|\hat{p}\right|$. 
In fact, $(\left|\hat{p}\right|-1)$ SWAP operations are required to move the control and target qubits of $g$ towards each other.
If none of the arrows of the path $\hat{p}$ on the coupling map (representing that a CNOT can be applied) points into the desired direction, we have to increase the true minimum costs further by 4, since 2 Hadamard operations are required before and after the CNOT to change its direction.

The heuristic costs of a mapping $\sigma^i_j$ can be determined from the real costs of each CNOT gate $g \in l_i$ in layer $l_i$. 
Simply summing them up might overestimate the true cost, because one SWAP operation might reduce the distance of the control and target qubits for more than one CNOT of layer~$l_i$. Since this would prevent us from determining the optimal solution~$\hat{\sigma}^i$, 
we instead determine the heuristic costs of a state~$\sigma^i_j$ as $h(\sigma^i_j) = \max_{g\in l_i} h(g,\sigma^i_j)$, i.e.~the maximum of the true costs of the CNOTs in layer $l_i$.

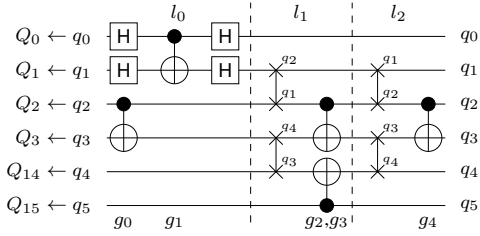
\begin{figure}
	\centering
	\scalebox{0.9}{
	\begin{tikzpicture}

	\draw[line width=0.300000] (0.500000,2.750000) -- (5.500000,2.750000);
	\draw (0.400000,2.750000) node [left] {$Q_0\leftarrow q_0$};
	\draw (5.600000,2.750000) node [right] {$q_0$};
	\draw[line width=0.300000] (0.500000,2.250000) -- (5.500000,2.250000);
	\draw (0.400000,2.250000) node [left] {$Q_1 \leftarrow q_1$};
	\draw (5.600000,2.250000) node [right] {$q_1$};
	\draw[line width=0.300000] (0.500000,1.750000) -- (5.500000,1.750000);
	\draw (0.400000,1.750000) node [left] {$Q_2\leftarrow q_2$};
	\draw (5.600000,1.750000) node [right] {$q_2$};
	\draw[line width=0.300000] (0.500000,1.250000) -- (5.500000,1.250000);
	\draw (0.400000,1.250000) node [left] {$Q_3 \leftarrow q_3$};
	\draw (5.600000,1.250000) node [right] {$q_3$};
	\draw[line width=0.300000] (0.500000,0.750000) -- (5.500000,0.750000);
	\draw (0.400000,0.750000) node [left] {$Q_{14}\leftarrow q_4$};
	\draw (5.600000,0.750000) node [right] {$q_4$};
	\draw[line width=0.300000] (0.500000,0.250000) -- (5.500000,0.250000);
	\draw (0.400000,0.250000) node [left] {$Q_{15}\leftarrow q_5$};
	\draw (5.600000,0.250000) node [right] {$q_5$};
		
	\draw[line width=0.3,fill=white] (0.55,2.55) rectangle ++(0.4,0.4); \node at (0.75,2.75) {\sf H};
	\draw[line width=0.3,fill=white] (0.55,2.05) rectangle ++(0.4,0.4); \node at (0.75,2.25) {\sf H};
	
	\draw[line width=0.300000] (0.75,1.05000) -- (0.75,1.750000);
	\draw[fill] (0.75,1.750000) circle (0.100000);
	\draw[line width=0.3] (0.75, 1.25) circle (0.2);
	
	\draw[line width=0.300000] (1.500,2.05000) -- (1.5000,2.750000);
	\draw[fill] (1.500,2.750000) circle (0.100000);
	\draw[line width=0.3] (1.5, 2.25) circle (0.2);
	
	\draw[line width=0.3,fill=white] (2.05,2.55) rectangle ++(0.4,0.4); \node at (2.25,2.75) {\sf H};
	\draw[line width=0.3,fill=white] (2.05,2.05) rectangle ++(0.4,0.4); \node at (2.25,2.25) {\sf H};
	
	\draw[line width=0.300000] (3.0000,1.750000) -- (3.0000,2.250000);
	\draw[line width=0.3] (2.9,2.15) -- ++(0.2,0.2) node[xshift=0.1cm, yshift=0.05cm] () {\tiny$q_2$} (2.9,2.35) -- ++(0.2,-0.2);
	\draw[line width=0.3] (2.9,1.65) -- ++(0.2,0.2) node[xshift=0.1cm, yshift=0.05cm] () {\tiny$q_1$} (2.9,1.85) -- ++(0.2,-0.2);
	
	\draw[line width=0.300000] (3.0000,0.750000) -- (3.0000,1.250000);
	\draw[line width=0.3] (2.9,1.15) -- ++(0.2,0.2) node[xshift=0.1cm, yshift=0.05cm] () {\tiny$q_4$} (2.9,1.35) -- ++(0.2,-0.2);
	\draw[line width=0.3] (2.9,0.65) -- ++(0.2,0.2) node[xshift=0.1cm, yshift=0.05cm] () {\tiny$q_3$} (2.9,0.85) -- ++(0.2,-0.2);
	
	\draw[line width=0.300000] (3.7500,1.05000) -- (3.75000,1.750000);
	\draw[fill] (3.7500,1.750000) circle (0.100000);
	\draw[line width=0.3] (3.75, 1.25) circle (0.2);
		
	\draw[line width=0.300000] (3.7500,0.25000) -- (3.75000,0.950000);
	\draw[fill] (3.7500,0.250000) circle (0.100000);
	\draw[line width=0.3] (3.75, 0.75) circle (0.2);
	
	\draw[line width=0.300000] (4.50000,1.750000) -- (4.50000,2.250000);
	\draw[line width=0.3] (4.4,2.15) -- ++(0.2,0.2) node[xshift=0.1cm, yshift=0.05cm] () {\tiny$q_1$} (4.4,2.35) -- ++(0.2,-0.2);
	\draw[line width=0.3] (4.4,1.65) -- ++(0.2,0.2) node[xshift=0.1cm, yshift=0.05cm] () {\tiny$q_2$} (4.4,1.85) -- ++(0.2,-0.2);

	\draw[line width=0.300000] (4.50000,0.750000) -- (4.50000,1.250000);
	\draw[line width=0.3] (4.4,1.15) -- ++(0.2,0.2) node[xshift=0.1cm, yshift=0.05cm] () {\tiny$q_3$} (4.4,1.35) -- ++(0.2,-0.2);
	\draw[line width=0.3] (4.4,0.65) -- ++(0.2,0.2) node[xshift=0.1cm, yshift=0.05cm] () {\tiny$q_4$} (4.4,0.85) -- ++(0.2,-0.2);
	
	\draw[line width=0.300000] (5.25,1.75000) -- (5.25,1.050000);
	\draw[fill] (5.25,1.750000) circle (0.100000);
	\draw[line width=0.3] (5.25, 1.25) circle (0.2);
	
	\draw[dashed] (2.625,0) -- (2.625,3.25);
	\draw[dashed] (4.125,0) -- (4.125,3.25);
	
	\node at (1.5625,3.1) {$l_0$};
	\node at (3.375,3.1) {$l_1$};
	\node at (4.8125,3.1) {$l_2$};

	\node at (0.75,0) {$g_0$};
	\node at (1.5,0) {$g_1$};
	\node at (3.75,0) {$g_2$,$g_3$};
	\node at (5.25,0) {$g_4$};	
	\end{tikzpicture}
}
	\caption{Circuit resulting from locally optimal mappings}
	\label{fig:circuit_locally_opt}
	
\end{figure}

\addtocounter{example}{-1}
\begin{example}[continued]
	The logical qubits $q_1$ and $q_4$ are mapped to the physical qubits $\sigma^1_0(q_1) = Q_1$ and \mbox{$\sigma^1_0(q_4) = Q_{14}$}, respectively. Since the shortest path on the coupling map  is $\hat{p} = Q_1 \rightarrow Q_2 \rightarrow Q_3 \rightarrow Q_{14}$ (cf.~Fig.~\ref{fig:coupling}), the true minimum costs for $g_2$ is $h(g_2, \sigma^1_0) = 2\cdot 7 = 14$.
	Analogously, the costs of $g_3$ can be determined to be $h(g_3, \sigma^1_0) = 7$ -- resulting in overall heuristic costs of $h(\sigma^1_0) = \max(14,7) =14$ for the initial mapping. 
	Following the A* algorithm outlined above, we eventually determine a mapping $\hat{\sigma}^1$ that maps the logical qubits $q_0$, $q_1$, $q_2$, $q_3$, $q_4$, and $q_5$ to the physical qubits $Q_0$, $Q_2$, $Q_1$, $Q_4$, $Q_3$, and $Q_5$ by inserting two SWAP operations (as depicted in Fig.~\ref{fig:circuit_locally_opt}). 
	Applying the algorithm also for mapping layer $l_2$, the circuit shown in Fig.~\ref{fig:circuit_locally_opt} results.
	This circuit is composed of 37 elementary operations and has depth~15. 
\end{example}

\subsection{Optimizations}
\label{sec:look_ahead}

$A^*$ allows us to efficiently determine an optimal mapping (by means of additionally required operations) for each layer.
However, the algorithm proposed in Section~\ref{sec:locally_opt} considers only a single layer when determining $\hat{\sigma}^i$ for layer $l_i$.

One way to optimize the proposed solution is to employ a look-ahead scheme which incorporates information from the following layers to the cost function.
To this end, we only have to change the heuristics to estimate the costs for reaching a mapping that satisfies all CNOT-constraints from the current one. In Section~\ref{sec:locally_opt}, we used the maximum of the costs for each CNOT gate in layer $l_i$ to estimate the true remaining cost. For the look-ahead scheme, we additionally determine an estimate for layer $l_{i+1}$. The overall heuristic that guides the search algorithm towards a solution is then the sum of both estimates. 

To incorporate the look-ahead scheme, we change the heuristics discussed in Section~\ref{sec:locally_opt}.
Instead of taking the maximum of the CNOTs in the current layer, we sum up the costs of all CNOTs in two layers (the current and the look-ahead layer), i.e.~$h(\sigma^i_j) = \sum_{g \in l_i \cup l_{i+1}} h(g, \sigma^i_j)$. As discussed above, this might lead to an over-estimation of the true remaining costs for reaching a goal state and, thus, the solution is not guaranteed to be locally optimal. However, this is not desired anyways, since we want to allow locally sub-optimal solutions in order to find cheaper mappings for the following layers -- resulting in smaller overall circuits.

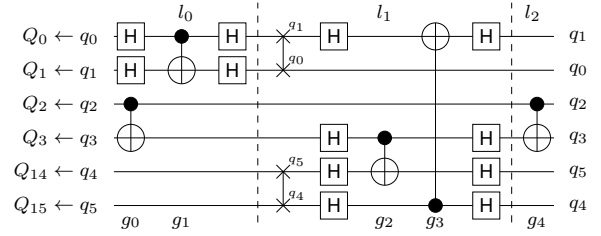
\begin{figure}
	\centering
	
	\scalebox{0.9}{
	\begin{tikzpicture}
	
	\draw[line width=0.300000] (0.500000,2.750000) -- (7.00000,2.750000);
	\draw (0.400000,2.750000) node [left] {$Q_0 \leftarrow q_0$};
	\draw (7.100000,2.750000) node [right] {$q_1$};
	\draw[line width=0.300000] (0.500000,2.250000) -- (7.00000,2.250000);
	\draw (0.400000,2.250000) node [left] {$Q_1 \leftarrow q_1$};
	\draw (7.100000,2.250000) node [right] {$q_0$};
	\draw[line width=0.300000] (0.500000,1.750000) -- (7.00000,1.750000);
	\draw (0.400000,1.750000) node [left] {$Q_2 \leftarrow q_2$};
	\draw (7.100000,1.750000) node [right] {$q_2$};
	\draw[line width=0.300000] (0.500000,1.250000) -- (7.00000,1.250000);
	\draw (0.400000,1.250000) node [left] {$Q_3 \leftarrow q_3$};
	\draw (7.100000,1.250000) node [right] {$q_3$};
	\draw[line width=0.300000] (0.500000,0.750000) -- (7.00000,0.750000);
	\draw (0.400000,0.750000) node [left] {$Q_{14} \leftarrow q_4$};
	\draw (7.100000,0.750000) node [right] {$q_5$};
	\draw[line width=0.300000] (0.500000,0.250000) -- (7.00000,0.250000);
	\draw (0.400000,0.250000) node [left] {$Q_{15} \leftarrow q_5$};
	\draw (7.100000,0.250000) node [right] {$q_4$};

	\draw[line width=0.3,fill=white] (0.55,2.55) rectangle ++(0.4,0.4); \node at (0.75,2.75) {\sf H};
	\draw[line width=0.3,fill=white] (0.55,2.05) rectangle ++(0.4,0.4); \node at (0.75,2.25) {\sf H};
	
	\draw[line width=0.300000] (0.75,1.05000) -- (0.75,1.750000);
	\draw[fill] (0.75,1.750000) circle (0.100000);
	\draw[line width=0.3] (0.75, 1.25) circle (0.2);
	
	\draw[line width=0.300000] (1.500,2.05000) -- (1.5000,2.750000);
	\draw[fill] (1.500,2.750000) circle (0.100000);
	\draw[line width=0.3] (1.5, 2.25) circle (0.2);
	
	\draw[line width=0.3,fill=white] (2.05,2.55) rectangle ++(0.4,0.4); \node at (2.25,2.75) {\sf H};
	\draw[line width=0.3,fill=white] (2.05,2.05) rectangle ++(0.4,0.4); \node at (2.25,2.25) {\sf H};
	
	\draw[line width=0.300000] (3.0000,2.750000) -- (3.0000,2.250000);
	\draw[line width=0.3] (2.9,2.65) -- ++(0.2,0.2) node[xshift=0.1cm, yshift=0.05cm] () {\tiny$q_1$} (2.9,2.85) -- ++(0.2,-0.2);
	\draw[line width=0.3] (2.9,2.15) -- ++(0.2,0.2) node[xshift=0.1cm, yshift=0.05cm] () {\tiny$q_0$} (2.9,2.35) -- ++(0.2,-0.2);
	
	\draw[line width=0.300000] (3.0000,0.750000) -- (3.0000,0.250000);
	\draw[line width=0.3] (2.9,0.65) -- ++(0.2,0.2) node[xshift=0.1cm, yshift=0.05cm] () {\tiny$q_5$} (2.9,0.85) -- ++(0.2,-0.2);
	\draw[line width=0.3] (2.9,0.15) -- ++(0.2,0.2) node[xshift=0.1cm, yshift=0.05cm] () {\tiny$q_4$} (2.9,0.35) -- ++(0.2,-0.2);
	
	\draw[line width=0.3,fill=white] (3.55,2.55) rectangle ++(0.4,0.4); \node at (3.75,2.75) {\sf H};
	\draw[line width=0.3,fill=white] (3.55,1.05) rectangle ++(0.4,0.4); \node at (3.75,1.25) {\sf H};
	\draw[line width=0.3,fill=white] (3.55,0.55) rectangle ++(0.4,0.4); \node at (3.75,0.75) {\sf H};
	\draw[line width=0.3,fill=white] (3.55,0.05) rectangle ++(0.4,0.4); \node at (3.75,0.25) {\sf H};
	
	\draw[line width=0.300000] (4.500,1.25000) -- (4.5000,0.550000);
	\draw[fill] (4.500,1.250000) circle (0.100000);
	\draw[line width=0.3] (4.5, 0.75) circle (0.2);
	
	\draw[line width=0.300000] (5.2500,0.25000) -- (5.25000,2.950000);
	\draw[fill] (5.2500,0.250000) circle (0.100000);
	\draw[line width=0.3] (5.25, 2.75) circle (0.2);
	
	\draw[line width=0.3,fill=white] (5.8,2.55) rectangle ++(0.4,0.4); \node at (6,2.75) {\sf H};
	\draw[line width=0.3,fill=white] (5.8,1.05) rectangle ++(0.4,0.4); \node at (6,1.25) {\sf H};
	\draw[line width=0.3,fill=white] (5.8,0.55) rectangle ++(0.4,0.4); \node at (6,0.75) {\sf H};
	\draw[line width=0.3,fill=white] (5.8,0.05) rectangle ++(0.4,0.4); \node at (6,0.25) {\sf H};
	
	\draw[line width=0.300000] (6.75,1.75000) -- (6.75,1.050000);
	\draw[fill] (6.75,1.750000) circle (0.100000);
	\draw[line width=0.3] (6.75, 1.25) circle (0.2);
	
	\draw[dashed] (2.625,0) -- (2.625,3.25);
	\draw[dashed] (6.375,0) -- (6.375,3.25);
	
	\node at (1.5675	,3.1) {$l_0$};
	\node at (4.5,3.1) {$l_1$};
	\node at (6.6875,3.1) {$l_2$};
	
	\node at (0.75,0) {$g_0$};
	\node at (1.5,0) {$g_1$};
	\node at (4.5,0) {$g_2$};
	\node at (5.25,0) {$g_3$};
	\node at (6.75,0) {$g_4$};

	\end{tikzpicture}
}
	\caption{Circuit generated when using the look-ahead scheme}
	\label{fig:look_ahead}
\end{figure}

\begin{example}
Consider again the quantum circuit shown in Fig.~\ref{fig:circuit1} and assume that the logical qubits $q_0$, $q_1$, $q_2$, $q_3$, $q_4$, and $q_5$ are mapped to the physical qubits $Q_0$, $Q_1$, $Q_2$, $Q_3$, $Q_{14}$, and $Q_{15}$, respectively. Using the look-ahead scheme discussed above will not determine the locally optimal solution with costs of 14 for layer $l_1$ (as discussed in Example~\ref{ex:locally_opt}), but a mapping $\hat{\sigma}^1$ that satisfies all CNOT-constraints with costs of 22 (as show in~Fig.~\ref{fig:look_ahead}).
The additional costs of 8 result since, after applying two SWAP gates (cf.~Fig.~\ref{fig:look_ahead}), the directions of both CNOTs of layer~$l_1$ have to change. However, this mapping also satisfies all \mbox{CNOT-constraints} for layer $l_2$, which means that the remaining CNOT $g_4$ can be applied without adding further SWAPs. The resulting  circuit is composed of a total of 31 elementary operations and has depth of 12 (as shown in Fig.~\ref{fig:look_ahead}; gates $g_2$ and $g_3$ can be applied concurrently). Consequently, the look-ahead scheme results in a cheaper mapping than the ``pure'' methodology proposed in Section~\ref{sec:locally_opt}
and yielding the circuit shown in Fig.~\ref{fig:circuit_locally_opt}.\footnote{Note that the graphical representation seems to be larger in Fig.~\ref{fig:look_ahead}. However, this is caused by the fact that the SWAP operations are not decomposed (cf.~Fig~\ref{fig:swap_decomposition}) in order to maintain readability.} 
\end{example}

Besides the look-ahead scheme, we can further improve the methodology  by not starting with a random mapping for layer~$l_0$. Instead, we propose to use partial mappings $\sigma_j^i$ and to start with an empty mapping $\sigma^0_0$ (i.e.~none of the logical qubits is mapped to a physical one).
Then, before we start to search a mapping for layer $l_1$, we check whether the qubits that occur in the CNOTs $g\in l_i$ have already been mapped for one of the former layers. If not, we can freely chose one of the ``free'' physical qubits (i.e.~a physical qubit no logical qubit is mapped to). Obviously, we choose the physical qubit so that the cost for finding $\hat{\sigma}^i$ is as small as possible.

This scheme gives us the freedom to evolve the initial mapping throughout the mapping process, rather than starting with an initial mapping that might be non-beneficial with respect to the overall number of elementary operations.

\begin{example}
	Optimizing the methodology with a partial mapping that is initially empty results in the circuit already shown before in Fig.~\ref{fig:circuit3}. This circuit is composed of 23 elementary operations and has depth 10 (gates $g_2$ and $g_3$ can be applied concurrently). 
\end{example}

\section{Experimental Evaluation}
\label{sec:results}

Taking all considerations and methods discussed above into account led to the development of a mapping methodology which decomposes arbitrary quantum functionality into elementary quantum gates supported by the QX architectures and, afterwards, maps them so that all CNOT-constraints are satisfied.
As mentioned above, IBM's Python SDK \emph{QISKit} already implements most of these steps, but lacks an efficient methodology for mapping the circuits such that all \mbox{CNOT-constraints} are satisfied. To overcome this issue, we have implemented the mapping methodology presented in this paper in C++ and integrated it into \emph{QISKit}. 
The adapted version of \emph{QISKit} as well as a standalone version of the methodology are publicly available at~\url{http://iic.jku.at/eda/research/ibm_qx_mapping}.

In this section, we compare the efficiency of the resulting scheme to the original design flow implemented in \emph{QISKit}~\cite{qiskit}. 
To this end, several functions taken from RevLib~\cite{WGT+:2008} as well as quantum algorithms written in Quipper~\cite{DBLP:conf/pldi/GreenLRSV13} or the Scaffold language~\cite{abhari2012scaffold} (and pre-compiled by the ScaffoldCC compiler~\cite{DBLP:conf/cf/JavadiAbhariPKHLCM14}) have been considered as benchmarks and mapped to the most recent 16-qubit architecture available (i.e.~\emph{QX5}).\footnote{We used all benchmarks that required at most 16 qubits since only these can be mapped to QX5.}
Besides that, benchmarks   
that are relevant for existing quantum algorithms such as 
quantum ripple-carry adders (based on the realization proposed in~\cite{CDKM:2005} and denoted \emph{adder}) and small versions of Shor's algorithm (based on the realization proposed in~\cite{DBLP:journals/qic/Beauregard03} and denoted~\emph{shor}) have been considered. 
All evaluations have been conducted on a 4.2\,GHz machine with 4 cores (2 hardware threads each) and 32\,GB~RAM.

\subsection{Effect of the Optimizations}
In a first series of evaluations, we evaluate the improvements gained by the optimizations discussed in Section~\ref{sec:look_ahead}. The corresponding numbers  are listed in Table~\ref{tab:results_parameters}. For each benchmark, we provide the name, the number of logical qubits~$n$, the number of gates $g$, as well as the depth of the circuit $d$, before mapping the circuit to the \emph{IBM QX5} architecture. 
In the remainder of the table, we list the results provided by the proposed methodology, i.e.~the number of gates $g$ and the depth of the circuit $d$ after mapping it to the \emph{IBM QX5} architecture as well as the time required to determine that mapping (in CPU seconds).

Three different settings of the methodology are thereby considered. As baseline serves the approach proposed in Section~\ref{sec:proposed} that uses an $A^*$ algorithm to determine locally optimal mappings for each layer of the circuit (denoted \emph{Baseline} in the following). Furthermore, we list the numbers when enriching the baseline with a look-ahead scheme as discussed in Section~\ref{sec:look_ahead} (denoted \emph{Look-Ahead} in the following). Finally, we also list the resulting numbers for the fully optimized methodology that uses a look-ahead scheme and  additionally allows for evolving the mapping throughout the mapping process as discussed in Section~\ref{sec:look_ahead} (denoted \emph{Fully-Optimized} in the following). The timeout was set to one hour.

\begin{table*}[t]
	\caption{Effect of the Optimizations}
	
	\label{tab:results_parameters}
	\centering
	\scriptsize
	\setlength{\tabcolsep}{0.1cm}
	
	\begin{tabular}{lrrr||rrr||rrr||rrr}
		\multicolumn{4}{c||}{}  & \multicolumn{3}{c||}{\emph{Baseline}} & \multicolumn{3}{c||}{\emph{Look-Ahead}} & \multicolumn{3}{c}{\emph{Fully-Optimized}} \\
		Name & $n$ & $g$ & $d$ & $g$ & $d$ & $t$ & $g$ & $d$ & $t$ & $g$ & $d$ & $t$ \\ \hline
		\begin{filecontents*}{parameter_study.csv}
file,qubits,gates,layers,base,,,look ahead,,,optimized,,
adder_10,10,142,99,444,251,1.22,355,209,1.08,292,172,1.20
hwb9,10,207775,116199,743973,403199,1662.82,653249,374954,1437.33,655220,375105,1422.33
ising\_model\_10,10,480,70,235,41,4.63,235,41,4.58,251,47,4.14
max46,10,27126,14257,,,,86049,46692,185.19,84914,46270,185.82
mini\_alu,10,173,69,710,301,2.38,587,261,1.32,474,225,1.25
qft\_10,10,200,63,685,227,1.23,445,135,1.54,447,170,1.25
rd73,10,230,92,952,405,1.83,916,374,1.57,656,301,1.52
sqn,10,10223,5458,37781,19461,80.15,32099,17785,72.25,32095,17801,68.97
sym9,10,21504,12087,78388,43269,172.95,67290,38982,147.99,66637,38849,145.37
sys6-v0,10,215,75,962,383,1.96,794,301,1.50,613,250,1.36
urf3,10,125362,70702,517104,271754,1045.85,439268,239099,888.77,440509,239702,873.84
9symml,11,34881,19235,133813,70088,296.80,114179,63659,255.02,116508,64279,254.25
dc1,11,1914,1038,8310,4277,16.22,6024,3359,13.07,5946,3378,12.38
life,11,22445,12511,86075,45499,358.56,73020,41137,161.90,74632,41767,166.95
shor\_11,11,49295,30520,125825,70115,325.13,109574,60721,317.81,106322,58943,322.78
sym9,11,34881,19235,133813,70088,292.40,114179,63659,249.49,116508,64279,251.42
urf4,11,512064,264330,1926128,980191,4257.19,1653689,888594,3481.55,1650845,878249,3534.79
wim,11,986,514,3632,1914,7.60,3176,1712,6.54,2985,1711,6.30
z4,11,3073,1644,12041,6332,24.91,10002,5486,20.71,9717,5335,20.92
adder\_12,12,177,123,631,336,1.76,483,249,1.64,372,226,1.32
cm152a,12,1221,684,4254,2226,9.13,4039,2271,8.23,3738,2155,8.02
cycle10\_2,12,6050,3386,23991,12405,49.62,19513,10950,45.82,19857,11141,42.26
rd84,12,13658,7261,52508,26668,157.72,45509,24421,107.69,45497,24473,99.89
sqrt8,12,3009,1659,11921,6224,26.64,10166,5642,21.35,9744,5501,19.66
sym10,12,64283,35572,251731,130657,535.84,214881,118780,500.05,215569,118753,501.02
sym9,12,328,127,1436,608,2.54,1240,532,2.27,955,425,2.08
adr4,13,3439,1839,13475,6829,29.53,11245,6120,23.84,11301,6205,23.17
dist,13,38046,19694,147115,72929,323.20,125342,66590,334.67,125867,66318,291.90
gse\_10,13,390180,245614,863511,533279,2441.38,576399,401121,2263.71,520010,376695,2237.10
ising\_model\_13,13,633,71,313,41,6.08,313,41,6.09,329,47,5.11
plus63mod4096,13,128744,72246,529896,270734,1203.45,434900,242815,1006.16,439981,243861,1086.48
radd,13,3213,1781,11790,6387,25.35,10868,6088,23.67,10441,5872,22.00
rd53,13,275,124,1367,619,2.53,1044,457,1.97,942,469,1.93
root,13,17159,8835,67941,32854,327.20,56654,29846,120.01,57874,30068,120.82
shor\_13,13,98109,59350,259511,140923,656.43,229752,121093,783.74,224556,118536,640.55
squar5,13,1993,1049,7948,4069,16.35,6453,3470,13.29,6267,3448,12.96
410184,14,211,104,914,441,1.82,708,337,1.42,758,366,1.48
adder\_14,14,212,147,,,,,,,437,268,1.47
clip,14,33827,17879,135455,67312,322.36,,,,114336,60882,327.55
cm42a,14,1776,940,6473,3394,13.93,5572,3076,11.16,5431,3013,11.95
cm85a,14,11414,6374,46300,23662,185.98,37927,21215,464.90,37746,21189,242.80
plus127mod8192,14,330777,185853,,,,,,,1132251,626451,2481.95
plus63mod8192,14,187112,105142,773514,395379,1628.19,637137,355040,1364.63,640204,354076,1443.33
pm1,14,1776,940,6473,3394,13.62,5572,3076,11.14,5431,3013,11.10
sao2,14,38577,19563,155351,74524,330.62,,,,131002,66975,283.90
sym6,14,270,135,1101,547,2.33,1136,526,2.05,852,456,1.84
co14,15,17936,8570,80399,34658,331.89,62348,29831,176.55,63826,30366,133.71
dc2,15,9462,5242,36968,19306,83.96,31722,17559,95.81,30680,17269,72.53
ham15,15,8763,4819,32175,17379,79.80,27861,15668,61.70,28310,15891,68.75
misex1,15,4813,2676,17833,9621,38.63,15260,8810,33.18,15185,8729,33.11
rd84,15,343,110,1593,553,3.30,1337,441,2.81,971,353,2.23
square\_root,15,7630,3847,,,,,,,25212,13205,55.35
urf6,15,171840,93645,684701,353581,1456.37,,,,580295,313011,1436.16
adder\_16,16,247,171,,,,,,,515,319,1.72
alu2,16,28492,15176,118919,58105,244.83,,,,98166,51817,454.93
cnt3-5,16,485,209,1957,887,3.89,1488,725,2.98,1376,669,3.00
example2,16,28492,15176,118919,58105,246.00,,,,98166,51817,449.08
inc,16,10619,5863,41042,21614,86.91,34742,19431,74.13,34375,19176,72.85
ising\_model\_16,16,786,71,391,41,6.88,391,41,6.86,426,48,6.47
qft\_16,16,512,105,2193,589,69.04,1299,281,8.62,1341,404,16.43
\end{filecontents*}
\csvreader[
		respect all, 
		late after line=\\,
		late after last line=,
		]{parameter_study.csv}
		{1=\Name,2=\Qubits, 3=\Gates, 4=\Depth, 5=\GA, 6=\LA, 7=\TA, 8=\GB, 9=\LB, 10=\TB, 11=\GC, 12=\LC, 13=\TC}
		{\Name  & \Qubits & \optnum{\Gates}{--} & \optnum{\Depth}{--} & \optnum{\GA}{--} & \optnum{\LA}{--} & \optnum{\TA}{TO} & \optnum{\GB}{--} & \optnum{\LB}{--} & \optnum{\TB}{TO} & \optnum{\GC}{--} & \optnum{\LC}{--} & \optnum{\TC}{TO} }
	\end{tabular}
\\
	\raggedright{
		$n$: the number of qubits \hspace{0.7cm} $g$: the number of quantum gates (elementary operations) \hspace{0.7cm} $d$: depth of the quantum circuits \hspace{0.7cm} $t$: runtime of the algorithm\\
	\emph{Baseline}: the approach described in Sec.~\ref{sec:locally_opt}\hspace{0.7cm}
	\emph{Look-Ahead}: the approach described in Sec.~\ref{sec:locally_opt} enriched with the look-ahead scheme discussed in Sec.~\ref{sec:look_ahead}\\ \emph{Fully-Optimized}: the approach described in Sec.~\ref{sec:locally_opt} enriched with all optimizations discussed in Sec.~\ref{sec:look_ahead}
}\vspace*{-5mm}
\end{table*}

Table~\ref{tab:results_parameters} clearly shows the improvements that can be gained by applying the optimizations discussed in Section~\ref{sec:look_ahead}. On average, the number of gates of the mapped circuit decreases by 16.1\% when applying a look-ahead scheme as discussed in Section~\ref{sec:look_ahead}. For the depth of the circuit, we obtain similar improvements. Here, the number of layers reduces on average by 13.4\%. However, using the look-ahead scheme causes the mapping algorithm to time out in nine cases (instead of five cases for \emph{baseline}) -- leading to a less scalable solution. 
If we additionally allow to evolve the initial mapping of logical qubits to physical qubits throughout the mapping process instead of starting with a random mapping, we can overcome this scalability issue while obtaining mappings of similar quality. In fact, the average improvement regarding the number of gates and the depth of the circuits slightly increase to 
19.7\% and 
14.1\%, respectively (compared to \emph{Baseline}). 

Overall, the optimizations discussed in Section~\ref{sec:look_ahead} not only increase the scalability of the mapping algorithm outlined in Section~\ref{sec:locally_opt}, but -- as a positive side effect -- also reduce the size of the resulting circuit.

\subsection{Comparison to the State of the Art}
In a second series of evaluation, we compare the proposed mapping methodology to the solution provided by IBM via \emph{QISKit}. A fair comparison of both mapping solution is guaranteed since we incorporated the mapping algorithm discussed in this paper into \emph{QISKit}. Hence, the same decomposition schemes as well as the same post-mapping optimizations are applied in both cases.

Table~\ref{tab:results} lists the respectively obtained results. 
For each benchmark, we again list the name, the number of logical qubits~$n$, the number of gates $g$, and the depth $d$ of the quantum circuit before mapping it to the \emph{IBM QX5} architecture. In the remaining columns, we list the number of gates, the depth, and the runtime $t$ (in CPU seconds) for IBM's solution as well as for the solution proposed in this work. Since IBM's mapping algorithm searches for mappings that satisfy all CNOT-constraints randomly (guided by certain heuristics), we conducted the mapping procedure 5 times for each benchmark and list the obtained minimum, the average (denoted by subscripts $_{min}$ and $_{avg}$, respectively), as well as the standard deviation $\sigma$ for each of the listed metrics.  
The timeout for searching a single mapping was again set to one hour.

The results clearly show that the proposed solution can efficiently tackle the considered mapping problem -- in particular compared to the method available thus far. 
While IBM's solution runs into the timeout of one hour in 10 out of 60 cases, the proposed algorithm determines a mapping for each circuit within the given time limit.
Besides that, the approach is frequently magnitudes faster compared to IBM's solution. 

Besides efficiency, the proposed methodology for mapping a quantum circuit to the \emph{IBM QX} architectures also yields circuits with significantly fewer gates 
than the results determined by IBM's solution. In fact, the solution proposed in Section~\ref{sec:proposed} results on average in circuits with 24.0\% fewer gates and 18.3\% fewer depth on average compared to the minimum observed when runnings IBM's algorithm several times. Compared to the average results yield by IBM's solution, we obtain improvements of 27.5\% and 22.0\% for gate count and circuit depth, respectively.

\begin{table*}[t]
	\caption{Mapping to the \emph{IBM QX5} architecture}
	
	\label{tab:results}
	\centering
	\scriptsize
	\setlength{\tabcolsep}{0.08cm}
	
	\begin{tabular}{lrrr||rrr|rrr|rrr||rrr}
		\multicolumn{4}{c||}{}  & \multicolumn{9}{c||}{IBM's solution} & \multicolumn{3}{c}{Proposed approach} \\
		Name & $n$ & $g$ & $d$ & $g_{min}$ & $g_{avg}$ & $\sigma_{g}$ & $d_{min}$ & $d_{avg}$ & $\sigma_{d}$ & $t_{min}$ & $t_{avg}$ & $\sigma_{t}$ & $g$ & $d$ & $t$ 	\\ \hline
		\begin{filecontents*}{results.csv}
file,,,,MIN,MIN,MIN,AVG,AVG,AVG,STD,STD,STD,size,depth,t
adder_10,10,142,99,382,203,5.75,421.00,223.80,6.71,29.37,16.85,0.72,292,172,1.20
hwb9,10,207775,116199,,,,,,,,,,655220,375105,1422.33
ising\_model\_10,10,480,70,347,73,6.23,401.40,89.60,7.07,79.37,21.46,0.99,251,47,4.14
max46,10,27126,14257,105651,53397,1652.31,106357.80,53664.20,1672.40,453.99,138.67,16.15,84914,46270,185.82
mini\_alu,10,173,69,707,290,9.82,736.00,303.40,10.32,19.37,10.78,0.42,474,225,1.25
qft\_10,10,200,63,670,210,9.49,755.00,241.40,10.90,98.89,30.16,2.35,447,170,1.25
rd73,10,230,92,930,393,12.74,1035.00,424.20,14.88,70.69,26.18,1.46,656,301,1.52
sqn,10,10223,5458,39175,20329,627.32,39563.60,20415.40,633.67,301.56,75.50,5.51,32095,17801,68.97
sym9,10,21504,12087,80867,43707,1314.22,81893.00,44128.40,1321.35,638.00,342.23,5.48,66637,38849,145.37
sys6-v0,10,215,75,853,329,11.29,945.80,346.20,12.80,51.52,12.22,1.17,613,250,1.36
urf3,10,125362,70702,,,,,,,,,,440509,239702,873.84
9symml,11,34881,19235,143042,73363,2233.67,144299.20,73821.00,2257.31,798.04,496.98,18.51,116508,64279,254.25
dc1,11,1914,1038,7283,3859,113.8,7356.40,3890.60,116.75,73.48,28.10,2.83,5946,3378,12.38
life,11,22445,12511,91724,47471,1446.67,92470.20,47860.80,1454.98,599.36,382.84,5.38,74632,41767,166.95
shor\_11,11,49295,30520,124160,67962,2149.13,125245.00,68367.40,2160.12,775.01,317.02,6.96,106322,58943,322.78
sym9,11,34881,19235,142431,72959,2237.2,143685.00,73595.40,2261.74,751.22,450.98,13.68,116508,64279,251.42
urf4,11,512064,264330,,,,,,,,,,1650845,878249,3534.79
wim,11,986,514,3834,1947,59.01,3897.20,1988.60,59.88,46.70,26.04,1.04,2985,1711,6.30
z4,11,3073,1644,11905,6024,188.53,12148.20,6190.80,192.41,134.51,87.10,3.74,9717,5335,20.92
adder\_12,12,177,123,579,279,8.81,636.00,307.00,10.28,68.67,26.10,1.14,372,226,1.32
cm152a,12,1221,684,4761,2501,74.61,4928.20,2581.20,76.53,100.23,64.31,1.18,3738,2155,8.02
cycle10\_2,12,6050,3386,25362,13125,392.12,25666.60,13224.00,394.36,301.28,80.01,2.43,19857,11141,42.26
rd84,12,13658,7261,56134,28172,860.9,56865.60,28393.40,874.21,425.85,126.79,10.25,45497,24473,99.89
sqrt8,12,3009,1659,12541,6398,194.8,12678.20,6457.60,197.02,127.15,52.37,1.36,9744,5501,19.66
sym10,12,64283,35572,,,,,,,,,,215569,118753,501.02
sym9,12,328,127,1411,582,20.09,1512.20,598.60,21.59,83.10,18.13,1.98,955,425,2.08
adr4,13,3439,1839,13638,6991,210.34,13958.80,7075.40,217.16,172.30,57.82,3.95,11301,6205,23.17
dist,13,38046,19694,158516,77027,2412.78,159655.00,77739.40,2445.98,1268.06,631.57,18.58,125867,66318,291.90
gse\_10,13,390180,245614,,,,,,,,,,520010,376695,2237.10
ising\_model\_13,13,633,71,439,82,7.87,573.80,138.20,9.53,101.73,44.62,1.37,329,47,5.11
plus63mod4096,13,128744,72246,,,,,,,,,,439981,243861,1086.48
radd,13,3213,1781,12674,6716,206.15,13263.00,6907.00,211.40,331.68,124.21,4.83,10441,5872,22.00
rd53,13,275,124,1223,518,16.65,1295.20,543.80,17.89,53.56,20.07,0.75,942,469,1.93
root,13,17159,8835,71721,34798,1094.63,72252.80,35009.00,1099.08,304.63,205.73,5.15,57874,30068,120.82
shor\_13,13,98109,59350,,,,,,,,,,224556,118536,640.55
squar5,13,1993,1049,8111,4073,124.09,8300.00,4132.20,125.77,201.58,59.48,1.47,6267,3448,12.96
410184,14,211,104,864,393,13.36,928.20,408.80,13.91,40.30,17.65,0.43,758,366,1.48
adder\_14,14,212,147,659,332,9.76,745.20,373.20,11.55,114.49,51.39,1.49,437,268,1.47
clip,14,33827,17879,144737,70732,2197.11,145459.00,71177.00,2212.02,496.53,291.36,12.55,114336,60882,327.55
cm42a,14,1776,940,6623,3480,104.23,6830.20,3538.20,109.00,170.16,52.18,2.47,5431,3013,11.95
cm85a,14,11414,6374,47908,24798,742.46,48885.40,25101.40,757.59,673.07,253.09,10.50,37746,21189,242.80
plus127mod8192,14,330777,185853,,,,,,,,,,1132251,626451,2481.95
plus63mod8192,14,187112,105142,,,,,,,,,,640204,354076,1443.33
pm1,14,1776,940,6488,3444,104.21,6809.60,3525.80,106.31,190.40,53.01,1.69,5431,3013,11.10
sao2,14,38577,19563,163679,77525,2495.54,164561.40,77771.40,2509.16,803.87,228.58,15.15,131002,66975,283.90
sym6,14,270,135,1092,514,16.05,1246.60,568.00,18.61,96.11,35.64,1.81,852,456,1.84
co14,15,17936,8570,83301,35926,1177.71,83649.40,36046.20,1192.44,234.76,98.35,8.36,63826,30366,133.71
dc2,15,9462,5242,38807,20155,601.92,39694.60,20359.20,625.13,602.91,148.41,16.72,30680,17269,72.53
ham15,15,8763,4819,35150,18293,546.05,35402.20,18453.20,552.47,273.69,116.27,5.29,28310,15891,68.75
misex1,15,4813,2676,19090,10172,299.9,19316.80,10235.80,304.52,169.21,60.85,3.18,15185,8729,33.11
rd84,15,343,110,1579,529,19.5,1807.60,599.20,23.18,159.89,46.33,2.47,971,353,2.23
square\_root,15,7630,3847,30349,14828,461.14,30760.80,15029.40,468.59,421.20,213.97,5.71,25212,13205,55.35
urf6,15,171840,93645,,,,,,,,,,580295,313011,1436.16
adder\_16,16,247,171,968,437,13.88,1039.40,473.60,14.94,51.79,29.18,1.01,515,319,1.72
alu2,16,28492,15176,125601,60839,1905.14,126758.20,61383.20,1922.29,906.96,386.58,15.33,98166,51817,454.93
cnt3-5,16,485,209,1899,825,27.08,2023.00,863.40,29.71,102.11,34.92,2.78,1376,669,3.00
example2,16,28492,15176,125022,60543,1900.17,127074.60,61347.20,1920.29,1176.85,440.11,13.42,98166,51817,449.08
inc,16,10619,5863,43097,22413,672.39,43561.20,22577.00,679.82,442.97,127.65,6.28,34375,19176,72.85
ising\_model\_16,16,786,71,785,149,12.44,858.00,172.40,13.64,51.98,16.84,0.79,426,48,6.47
qft\_16,16,512,105,2056,521,25.6,2219.00,542.60,27.10,128.72,16.27,1.76,1341,404,16.43

\end{filecontents*}
\csvreader[
		respect all, 
		late after line=\\,
		late after last line=,
		]{results.csv}
		{1=\Name,2=\Qubits, 3=\Gates, 4=\Depth, 5=\gMin, 6=\dMin, 7=\tMin, 8=\gAvg, 9=\dAvg, 10=\tAvg, 11=\gStd, 12=\dStd, 13=\tStd, 14=\gProp, 15=\dProp, 16=\tProp}
		{\Name  & \Qubits & \optnum{\Gates}{--} & \optnum{\Depth}{--} & \optnum{\gMin}{--} & \optnum{\gAvg}{--} & \optnum{\gStd}{--} & \optnum{\dMin}{--} & \optnum{\dAvg}{--} & \optnum{\dStd}{--} & \optnum{\tMin}{TO} & \optnum{\tAvg}{--} & \optnum{\tStd}{--} & \optnum{\gProp}{--} & \optnum{\dProp}{--} & \optnum{\tProp}{TO} }
	\end{tabular}
\\
\raggedright{
	$n$: the number of qubits \hspace{0.7cm} $g$: the number of quantum gates (elementary operations) \hspace{0.7cm} $d$: depth of the quantum circuits \hspace{0.7cm} $t$: runtime of the algorithm\\
	For IBM's solution, we list the obtained minimum, the average, and the standard deviation of 5 runs (denoted by $_{min}$, $_{avg}$, and $\sigma$, respectively).	
}
\end{table*}

\section{Conclusions}\label{sec:conclusion}

In this paper, we proposed an advanced and integrated methodology
that efficiently maps a given quantum circuit to IBM's QX architectures.
To this end, the desired quantum functionality is first decomposed into the supported elementary quantum gates. Afterwards, CNOT-constraints imposed by the architecture are satisfied. Particular the later step caused a non-trivial task for which an efficient solution based on a depth-based partitioning, an $A^*$ search algorithm, a look-ahead scheme, as well as a dedicated initialization of the mapping has been proposed. The resulting approach eventually allows us to efficiently map quantum circuits to real quantum hardware and has been integrated into IBM's SDK \emph{QISKit}. The efficiency has been confirmed by experimental evaluations. The proposed approach was able to determine a mapping for quantum circuits within seconds in most cases whereas IBM's solution requires more than one hour to determine a solution for several cases. As a further positive side effect, the  mapped circuits have significantly fewer gates and smaller circuit depth, which positively influences the reliability and the runtime of the circuit.
The resulting methodology is generic, i.e.~it can be directly applied to all existing QX architectures as well as similar architectures which may come in the future. All implementations are publicly available at~\url{http://iic.jku.at/eda/research/ibm_qx_mapping}.

\section*{Acknowledgments}

This work has partially been supported by the European Union through the COST Action IC1405 and the Linz Institute of Technology (CHARON).

\bibliographystyle{ieeetr}
{
	\bibliography{lit_mapping}
}

\begin{IEEEbiography}
	[{\includegraphics[width=1in,height=1.25in,clip,keepaspectratio]{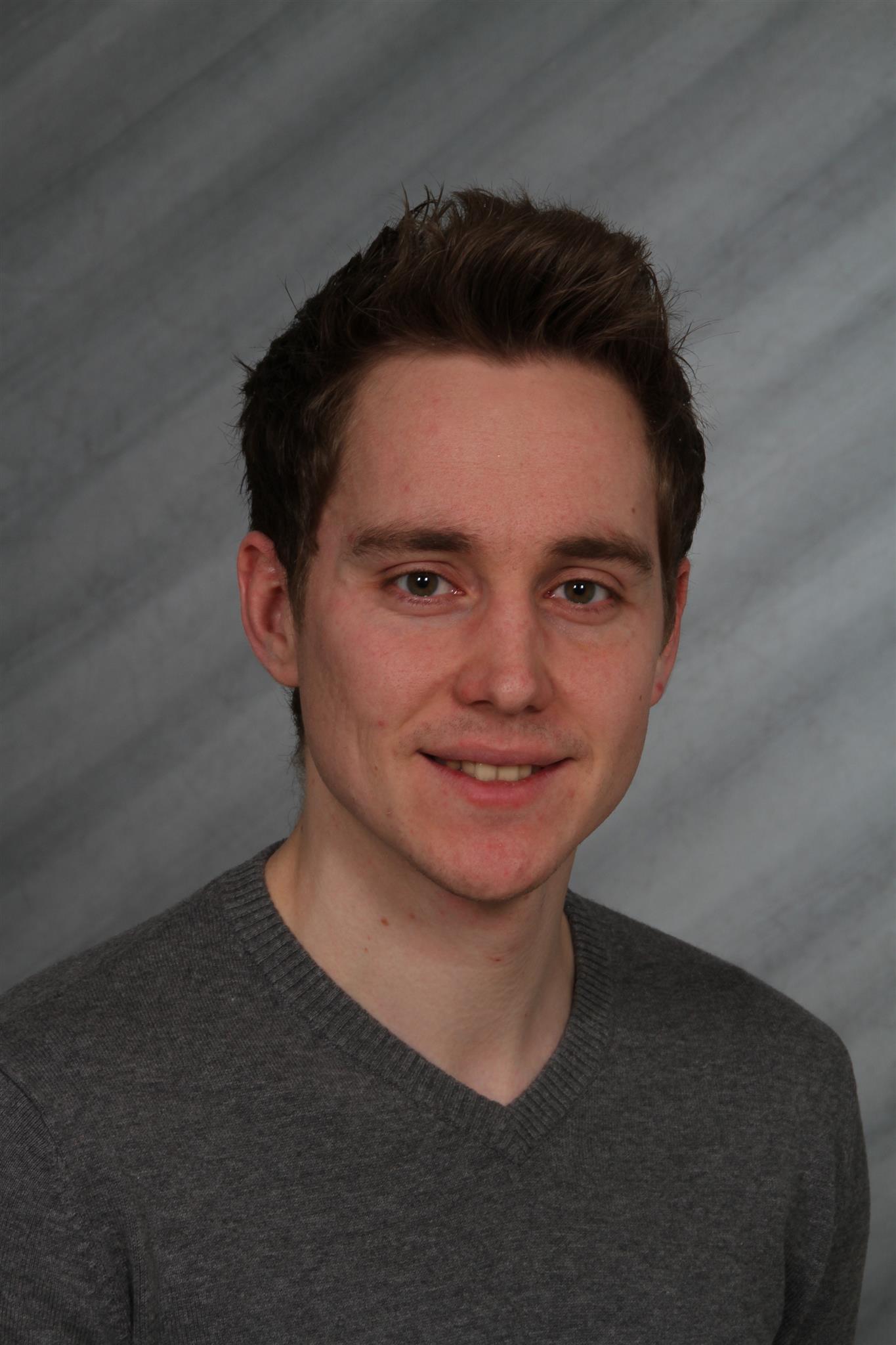}}]{Alwin Zulehner}
Alwin Zulehner (S’17) received his BSc and MSc degree in computer science from the Johannes Kepler University Linz, Austria in 2012 and 2015, respectively.
He is currently a Ph.D. student at the Institute for Integrated Circuits at the Johannes Kepler University Linz, Austria. 
His research interests include design automation for emerging technologies, currently focusing on reversible circuits and quantum circuits. In these areas, he has published several papers on international conferences and journals such as the IEEE Transactions on Computer Aided Design of Integrated Circuits and Systems (TCAD), Asia and South Pacific Design Automation Conference (ASP-DAC), Design, Automation and Test in Europe (DATE) and International Conference on Reversible Computation.
\end{IEEEbiography}

\begin{IEEEbiography}
	[{\includegraphics[width=1in,height=1.25in,clip,keepaspectratio]{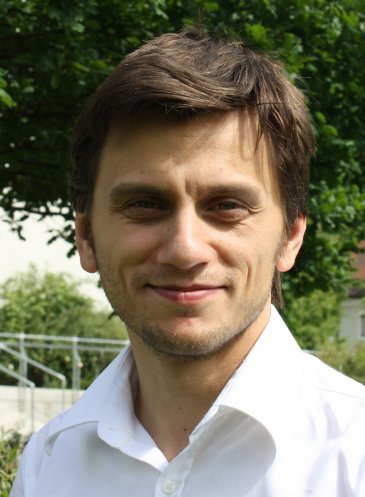}}]{Alexandru Paler}
Alexandru Paler is a postdoc at the Johannes Kepler University Linz. His research focuses on algorithms for compiling and assembling large scale error corrected quantum circuits.
\end{IEEEbiography}

\begin{IEEEbiography}
	[{\includegraphics[width=1in,height=1.25in,clip,keepaspectratio]{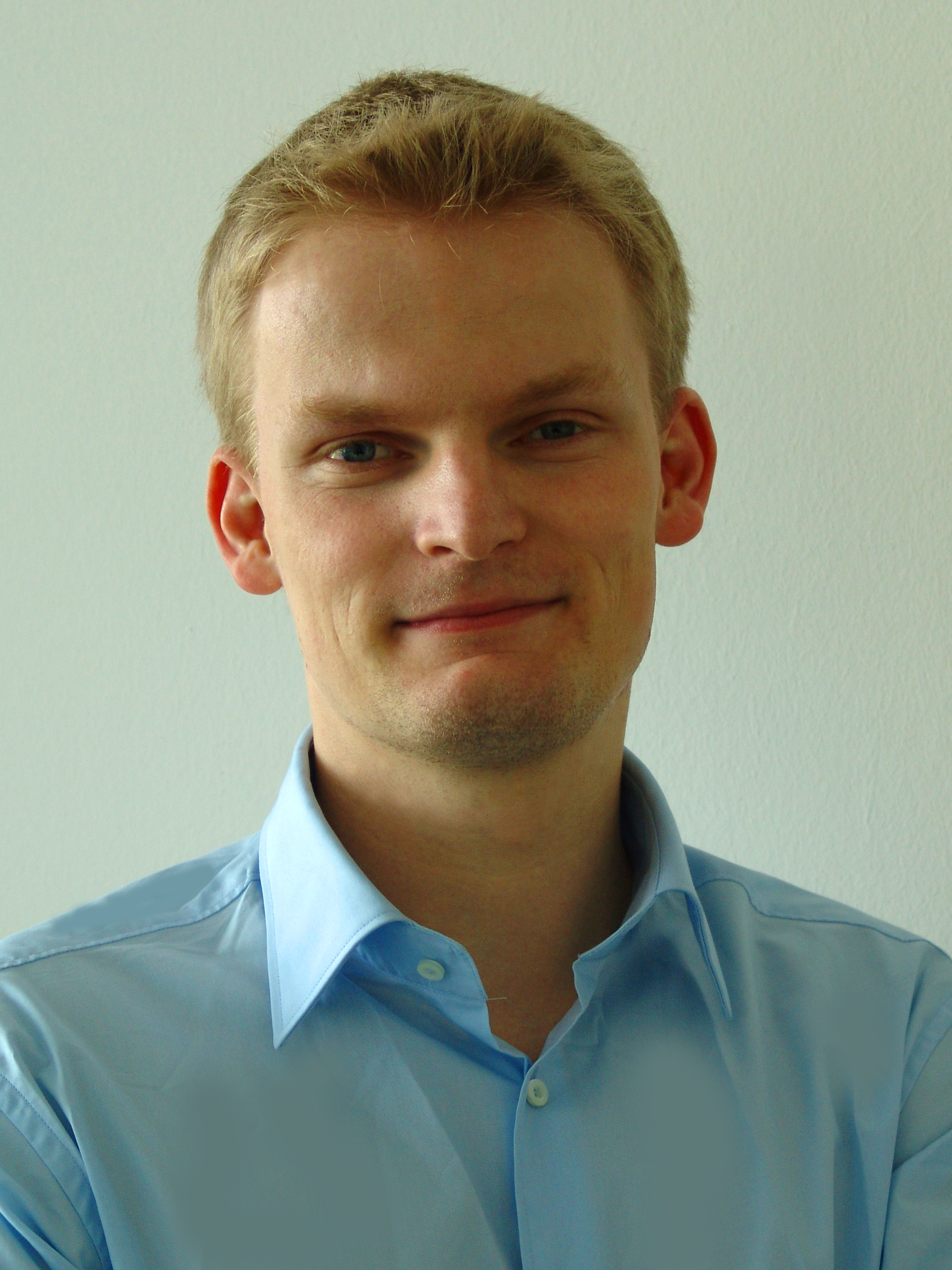}}]{Robert Wille}
Robert Wille (M’06–SM’09) received the Diploma and Dr.-Ing. degrees in computer science from the University of Bremen, Bremen, Germany, in 2006 and 2009, respectively. He was with the Group of Computer Architecture, University of Bremen, from 2006 to 2015, and has been with the German Research Center for Artificial Intelligence (DFKI), Bremen, since 2013. He was a Lecturer with the University of Applied Science of Bremen, Bremen, Germany, and a Visiting Professor with the University of Potsdam, Potsdam, Germany, and Technical University Dresden, Dresden, Germany. Since 2015, he is a Full Professor with Johannes Kepler University Linz, Linz, Austria. His current research interests include the design of circuits and systems for both conventional and emerging technologies. In these areas, he has published over 200 papers in journals and conferences. Dr. Wille has served in Editorial Boards and Program Committees of numerous journals/conferences, such as the IEEE Transactions on Computer Aided Design of Integrated Circuits and Systems  (TCAD), Asia and South Pacific Design Automation Conference (ASP-DAC), Design Automation Conference (DAC), Design, Automation and Test in Europe (DATE) and International Conference on Computer Aided Design (ICCAD).
\end{IEEEbiography}

\end{document}